\documentclass[twocolumn,pra,superscriptaddress,showpacs,amsmath]{revtex4}

\usepackage{graphicx}

\begin{document}

\title
     {{\it REVIEW}
     
     Quantum optics  with ultracold quantum gases:

     towards the full quantum regime of the light-matter interaction}
\date{\today}

\author{Igor B. Mekhov}
\affiliation{University of Oxford, Department of Physics, Clarendon Laboratory, Parks Road, Oxford OX1 3PU, UK}
\author{Helmut Ritsch}
\affiliation{Institut f\"ur Theoretische Physik, Universit\"at Innsbruck, Technikerstr. 25, 6020 Innsbruck, Austria}

\begin{abstract}
Although the study of ultracold quantum gases trapped by light is a prominent direction of modern research, the quantum properties of light were widely neglected in this field. Quantum optics with quantum gases closes this gap and addresses phenomena, where the quantum statistical nature of both light and ultracold matter play equally important roles. First,  light  can serve as a quantum nondemolition (QND) probe of the quantum dynamics of various ultracold particles from ultracold atomic and molecular gases to nanoparticles and nanomechanical systems. Second, due to dynamic light-matter entanglement, projective measurement-based preparation of the many-body  states is possible, where the class of emerging atomic states can be designed via optical geometry. Light scattering constitutes such a quantum measurement with controllable measurement back-action. As in cavity-based spin squeezing, atom number squeezed and Schr{\"o}dinger cat states can be prepared. Third, trapping atoms inside an optical cavity one creates optical potentials and forces, which are not prescribed but quantized and  dynamical variables themselves. Ultimately,  cavity QED with quantum gases requires a self-consistent solution for light and particles, which enriches the picture of quantum many-body states of atoms trapped in quantum potentials. This will allow quantum simulations of phenomena related to the physics of phonons, polarons, polaritons and other quantum quasiparticles.
\end{abstract}


\maketitle

\section{Introduction}

Both quantum optics with nonclassical light and physics of ultracold quantum gases are  well-established and active fields at the forefront of modern quantum physics\cite{BlochDalibard,Lewenstein}. However, the
integration of the two fields is far from being complete.

Historically, optics treating the light as electromagnetic waves is one of the most developed and
fruitful fields of physics. It has provided a lot of technological breakthroughs and the highest level of measurement
precision. Quantum optics extends this to a proper description of the quantum fluctuations of light beyond a mean-field
description. Today it is also a well-established field, both theoretically and experimentally
\cite{Scully}.

The progress in laser cooling techniques in the last decades of the
20th century led to the foundation of a new field of atom physics:
atom optics. It was shown that the matter waves of ultracold atoms
can be treated similar to light waves in classical optics and can be
manipulated using laser light forces.
The quantum properties of matter waves beyond the mean-field
description became apparent after 1995, when the first
Bose-Einstein condensate (BEC) and many other fascination quantum
states of bosonic and fermionic ultracold atoms were obtained
\cite{BlochDalibard,Lewenstein}. An exciting demonstration of
"quantum atom optics" was presented in 2002, when the quantum phase
transition between two states of atoms with nearly the same mean
density, but radically different quantum fluctuations was obtained, i.e., the
superfluid (SF) to Mott insulator (MI) state transition
\cite{Jaksch,BlochSFMI}.

The roles of light and matter in optics and atom optics are
completely reversed. Various devices known in optics as
beam-splitters, mirrors, diffraction grating, etc. are created using
light forces and applied to matter waves. However, up to now, the
absolute majority of even very involved setups and theoretical
models in physics of ultracold quantum gases treat light as an essentially
classical auxiliary tool to prepare and probe intriguing atomic
states. In this context, the periodic micropotentials of light
(optical lattices) play the role of cavities in optics enabling one to
store and manipulate various atomic quantum states.

Quantum optics of quantum gases should close the gap between
quantum optics and atom optics by addressing phenomena, where the
quantum natures of both light and matter play equally important
roles. Experimentally, such an ultimate quantum level of the
light-matter interaction became feasible only recently, when the
quantum gas was coupled to the mode of a high-Q cavity
\cite{Brennecke,Colombe,Slama}. Even scattering of quantized light from a BEC in free (without a cavity) space was not realized so far. However, implementing cavity quantum electrodynamics (QED) setups with quantum gases as recently achieved
provides for the most interesting and controllable implementation to study the atom- and light-generated
quantum effects.

On the one hand, the quantum properties of atoms will manifest
themselves in the scattered light, which will lead to novel
nondestructive methods of probing and manipulating atomic states by
light measurement. On the other hand, the quantization of light (i.e. trapping
potentials) will modify atomic many-body dynamics well-known only for
classical potentials and give rise to novel quantum phases.

This paper is organized as follows. In Sec. II, we formulate a general theoretical model, which treats the light and atomic motion in a fully quantum way. As a particular example, a generalized Bose-Hubbard model will be derived for quantized optical lattice potentials. Section III considers quantum non-demolition (QND) setups for probing the quantum states of ultracold particles by light scattering. In this section, we establish the relations between various light- and matter-related observables (such as expectation values of different quantities, their distribution functions and correlation functions). In Sec. IV, we go beyond the standard goal of the determination of correlation and distribution functions: we consider the truly quantum nature of the measurement process itself, which is evident in a single run of an optical measurement without averaging over many realizations. The quantum measurement back-action is used as an active tool to prepare various many-body states of ultracold particles. In Sec. V, we proceed deeper into the quantum regime and analyze the situation, where not only the quantization of probe light and atomic motion are important, but even the quantum nature of the trapping potential is crucial. This brings us to the formulation of the concept for "quantum optical lattices" produced inside a high-Q cavity. Various applications of the cavity QED with quantum gases are mentioned in Sec. VI, and conclusions are presented in Sec. VII.

\section{Theoretical model}

\begin{figure}
\scalebox{0.55}[0.55]{\includegraphics{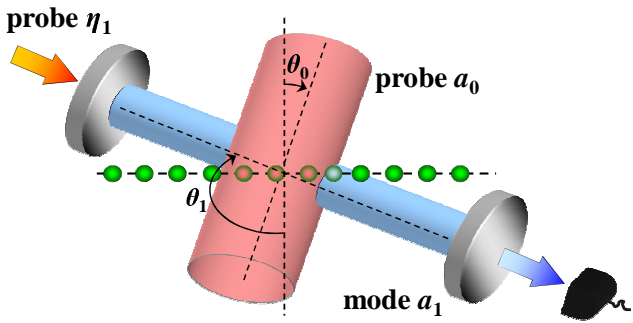}}
\caption{\label{fig1} Setup. The atoms periodically trapped in a lattice are illuminated
by the transverse probe $a_0$ (the trapping laser beams are not shown). The scattered light mode $a_1$ is collected by a cavity. Another probe $\eta_1$ through a cavity mirror can be present. The photons leaking the cavity are measured by the detector.}
\end{figure}

Here, we formulate the theoretical model to describe the interaction of an ultracold quantum gas with quantized light \cite{ICAP06,PRL07,PRA07,EPJD08}. The spatial geometry of light is taken into account in a very general form: the model can describe the interaction inside a cavity or in free space. The trapping of atoms by fully quantum potential is captured by this model as well. Describing the quantum properties of atoms, we focus on a particular case: spinless bosons. Generalizations to fermions and spin particles is straightforward and of interest as well.

We consider an ensemble of $N$ two-level atoms in an optical lattice with $M$ sites. In quantum optics of quantum gases, a restriction to optical lattices is very convenient as it allows to precisely describe the many-body atomic state for a broad range of parameters. Moreover, experimentally different setups can be described by the same general model. The typical examples include a multi-site lattice with the low filling factor (e.g. one or two atoms per lattice site as in a typical superfluid to Mott insulator transition setup) and a trapping potential with the high filling factors but small number of sites (e.g. a BEC in a double-well potential setup).

In general, the atoms are trapped in a lattice potential created by strong lasers as it is usual in the standard cold atom problems. In the presence of such a classical potential,
atoms are illuminated and scatter light at different
directions. As shown in Fig.~\ref{fig1}, the atoms in a lattice (green) are illuminated by a probe beam (red), and the measurements are carried out in the direction of one of a scattered light modes (blue). In fact, the probe and scattered light modes can be in free space without the presence of any cavity. In practice, light modes can be selected by traveling- or standing-wave cavities, or even correspond to different modes of the same cavity. One important reason for including a cavity is to enhance light scattering into some particular direction. Another reason, is that the cavity mode can form a fully quantum trapping potential for the atoms (even if no classical trapping potential is present). For definiteness, we will
consider the case, where the light mode functions are determined by cavities,
whose axes directions can be varied with respect to the lattice axis
(the simplest case of two waves, probe beam and cavity mode, at angles
$\theta_0$ and $\theta_1$ is shown in Fig.~\ref{fig1}). Instead of
varying the angles, also the light wavelengths can be varied with respect
to the wavelength of a trapping beam. We also assume, that not all
$M$ lattice sites are necessarily illuminated,
but only a subset of $K\le M$ sites. The selection of $K$ out of the total $M$ sites enriches the picture. In the simplest case, a continuous part of a lattice with $K$ sites can be illuminated. However, the nontrivial selection of the illuminated sites is also possible: e.g., each second site can be easily illuminated by choosing the light wavelength twice as the wavelength of the trapping beam. Moreover, using several probe beams one can make the optical geometry even more interesting.

A quite general model many-body Hamiltonian in the second quantized form is given by \cite{PRL07,PRA07,EPJD08}
\begin{subequations}\label{1}
\begin{eqnarray}
H=H_f +H_a, \\
H_f=\sum_l{\hbar\omega_l a^\dag_l a_l} -i\hbar\sum_l{(\eta^*_l a_l -
\eta_l a^\dag_l)}, \\
H_a=\int{d^3{\bf r}\Psi^\dag({\bf r})H_{a1}\Psi({\bf r})} \nonumber\\
+\frac{2\pi a_s \hbar^2}{m}\int{d^3{\bf r}\Psi^\dag({\bf
r})\Psi^\dag({\bf r})\Psi({\bf r})\Psi({\bf r})}.
\end{eqnarray}
\end{subequations}
In the field part of the Hamiltonian $H_f$, $a_l$ are the
annihilation operators of light modes with the frequencies
$\omega_l$, wave vectors ${\bf k}_l$, and mode functions $u_l({\bf
r})$, which can be pumped by coherent fields with amplitudes
$\eta_l$. In the atom part, $H_a$, $\Psi({\bf r})$ is the atomic
matter-field operator, $a_s$ is the $s$-wave scattering length
characterizing the direct interatomic interaction, and $H_{a1}$ is
the atomic part of the single-particle Hamiltonian $H_1$, which in
the rotating-wave and dipole approximation has a form

\begin{subequations}\label{2}
\begin{eqnarray}
H_1=H_f +H_{a1}, \\
H_{a1}=\frac{{\bf p}^2}{2m_a}+\frac{\hbar\omega_a}{2} \sigma_z -
i\hbar \sum_l{[\sigma^+ g_l a_l u_l({\bf r})-\text{H. c.}}]
\end{eqnarray}
\end{subequations}
Here, ${\bf p}$ and ${\bf r}$ are the momentum and position
operators of an atom of mass $m_a$ and resonance frequency
$\omega_a$, $\sigma^+$, $\sigma^-$, and $\sigma_z$ are the raising,
lowering, and population difference operators, $g_l$ are the
atom--light coupling constants for each mode. The inclusion of the interaction between the atom and quantum light in the single-particle Hamiltonian is the key step, which is different from the standard problems of ultracold atoms in classical potentials.

We will consider essentially nonresonant interaction where the
light-atom detunings $\Delta_{la} = \omega_l - \omega_a$ are much
larger than the spontaneous emission rate and Rabi frequencies $g_l
a_l$. Thus, in the Heisenberg equations obtained from the
single-atom Hamiltonian $H_1$ (\ref{2}), the atomic population difference $\sigma_z$ can be set to
$-1$ (approximation of linear dipoles, i.e., the dipoles responding linearly to the light amplitude with the negligible population of the excited state). Moreover, the polarization
$\sigma^-$ can be adiabatically eliminated and expressed via the
fields $a_l$. The elimination of $\sigma^-$ and setting $\sigma_z=-1$ is equivalent to the adiabatic elimination of the upper atomic level. The light-atom detunings can be all then replaced by a single value $\Delta_{a} = \omega_p - \omega_a$, where $\omega_p$ is, for example, the frequency of the external probe. An effective single-particle Hamiltonian that gives
the corresponding Heisenberg equation for $a_l$ can be written as
$H_{1\text{eff}}=H_f +H_{a1}$ with
\begin{eqnarray}\label{3}
H_{a1}=\frac{{\bf p}^2}{2m_a}+V_{\text {cl}}({\bf r})+\frac{\hbar}{\Delta_{a}}
\sum_{l,m}{u_l^*({\bf r})u_m({\bf r}) g_l g_m a^\dag_l
a_m}.
\end{eqnarray}
Here, we have also added a classical trapping potential of the
lattice, $V_{\text {cl}}({\bf r})$, which corresponds to a strong
classical standing wave. This potential can be, of course, derived
from one of the modes $a_l = a_{\text {cl}}$ [in this case $V_{\text
{cl}}({\bf r})=\hbar g^2_{\text {cl}}|a_{\text {cl}} u_{\text {cl}}({\bf
r})|^2/\Delta_{\text {cl}a}$], and it can scatter light into other
modes \cite{NimmrichterNJP2010}. Nevertheless, at this point we will consider $V_{\text
{cl}}({\bf r})$ as an independent potential, which does not affect
light scattering of other modes that will be significantly detuned
from $a_{\text {cl}}$ [i.e. the interference terms between $a_{\text
{cl}}$ and other modes are not considered in the last term of
Eq.~(\ref{3})]. A later inclusion of the light scattered by the
trapping wave will not constitute a difficulty, due to the linearity
of dipoles assumed in this model.

If one considers scattering of weak modes from the atoms in a deep
lattice, the fields $a_l$ are much weaker than the field
forming the potential $V_{\text {cl}}({\bf r})$. To derive the
generalized Bose--Hubbard Hamiltonian near zero temperature, we expand the field operator
$\Psi({\bf r})$ in Eq.~(\ref{1}), using localized Wannier functions
corresponding to $V_{\text {cl}}({\bf r})$ and keeping only the
lowest vibrational state at each site: $\Psi({\bf
r})=\sum_{i=1}^{M}{b_i w({\bf r}-{\bf r}_i)}$, where $b_i$ is the
annihilation operator of an atom at the site $i$ with the coordinate
${\bf r}_i$. Substituting this expansion in Eq.~(\ref{1}) with
$H_{a1}$ (\ref{3}), we get
\begin{eqnarray}\label{4}
H=H_f+\sum_{i,j=1}^M{J_{i,j}^{\text {cl}}b_i^\dag b_j} \nonumber \\
+ \frac{\hbar}{\Delta_{a}}
\sum_{l,m}{g_l g_m a^\dag_l
a_m}\left(\sum_{i,j=1}^K{J_{i,j}^{lm}b_i^\dag
b_j}\right)  \nonumber \\
+\frac{U}{2}\sum_{i=1}^M{b_i^\dag b_i(b_i^\dag b_i-1)},
\end{eqnarray}
where the coefficients $J_{ij}^{\text {cl}}$ correspond to the
quantum motion of atoms in the classical potential and are typical
for the Bose--Hubbard Hamiltonian \cite{Jaksch}:

\begin{equation}\label{5}
J_{i,j}^{\text {cl}}=\int{d{\bf r}}w({\bf r}-{\bf
r}_i)\left(-\frac{\hbar^2\nabla^2}{2m}+V_{\text {cl}}({\bf
r})\right)w({\bf r}-{\bf r}_j).
\end{equation}
However, in contrast to the usual Bose--Hubbard model, one has new
terms depending on the coefficients $J_{ij}^{lm}$, which describe an
additional contribution arising from the presence of light modes:
\begin{equation}\label{6}
J_{i,j}^{lm}=\int{d{\bf r}}w({\bf r}-{\bf r}_i) u_l^*({\bf
r})u_m({\bf r})  w({\bf r}-{\bf r}_j).
\end{equation}
In the last term of Eq.~(\ref{4}), only the on-site interaction was
taken into account and $U=4\pi a_s\hbar^2/m_a \int{d{\bf r}|w({\bf
r})|^4}$.

Note, that if the contribution of the quantized light is not much weaker than the contribution of the classical potential, or if the classical potential is not present at all, the Wannier functions should be determined in a self-consistent way taking into account the mean depth of the quantum potential, generated by the quantum light modes. This indeed significantly complicates the theoretical treatment \cite{EPJD08}. Although the form of the above equations still holds, the coefficients (\ref{5}) and (\ref{6}) will depend on the quantum state of light as well.

As a usual approximation, we restrict atom tunneling to the nearest neighbor sites. Thus, coefficients (\ref{5}) do
not depend on the site indices ($J_{i,i}^{\text {cl}}=J_0^{\text
{cl}}$ and $J_{i,i\pm 1}^{\text {cl}}=J^{\text {cl}}$), while
coefficients (\ref{6}) are still index-dependent. The Hamiltonian
(\ref{4}) then reads
\begin{eqnarray}\label{7}
H=H_f+J_0^{\text {cl}}\hat{N}+J^{\text {cl}}\hat{B} \nonumber \\
+\frac{\hbar}{\Delta_{a}}
\sum_{l,m}{g_l g_m a^\dag_l
a_m}\left(\sum_{i=1}^K{J_{i,i}^{lm}\hat{n}_i}\right)  \nonumber \\
+\frac{\hbar}{\Delta_{a}} \sum_{l,m}{g_l g_m a^\dag_l
a_m}\left(\sum_{<i,j>}^K{J_{i,j}^{lm}b_i^\dag
b_j}\right) \nonumber \\
+\frac{U}{2}\sum_{i=1}^M{\hat{n}_i(\hat{n}_i-1)},
\end{eqnarray}
where $<i,j>$ denotes the sum over neighboring pairs,
$\hat{n}_i=b_i^\dag b_i$ is the atom number operator at the $i$-th
site, and $\hat{B}=\sum_{i=1}^M{b^\dag_i b_{i+1}}+{\text {H.c.}}$
While the total atom number determined by
$\hat{N}=\sum_{i=1}^M{\hat{n}_i}$ is conserved, the atom number at
the illuminated sites, determined by
$\hat{N}_K=\sum_{i=1}^K{\hat{n}_i}$, is not necessarily a conserved
quantity.

The Heisenberg equations for $a_l$ and $b_i$ can be obtained from
the Hamiltonian (\ref{7}) as
\begin{subequations}\label{8}
\begin{eqnarray}
\dot{a}_l= -i\left( \omega_l
+\frac{g_l^2}{\Delta_{a}}\sum_{i=1}^K{J_{i,i}^{ll}\hat{n}_i}+
\frac{g_l^2}{\Delta_{a}}\sum_{<i,j>}^K{J_{i,j}^{ll}b_i^\dag
b_j}\right) a_l    \nonumber \\
-i\frac{g_l}{\Delta_{a}}\sum_{m \ne
l}g_m a_m\left(\sum_{i=1}^K{J_{i,i}^{lm}\hat{n}_i}\right) \nonumber \\
-i\frac{g_l}{\Delta_{a}}\sum_{m \ne
l}g_m a_m\left(\sum_{<i,j>}^K{J_{i,j}^{lm}b_i^\dag
b_j}\right)+ \eta_l -\kappa_l a_l, \label{8a}    \\
\dot{b}_i=-\frac{i}{\hbar}\left( J_0^{\text {cl}}+\frac{\hbar}{\Delta_{a}}
\sum_{l,m}{g_l g_m a^\dag_l a_m J_{i,i}^{lm}}+U\hat{n}_i\right) b_i  \nonumber \\
-\frac{i}{\hbar}\left( J^{\text {cl}}+\frac{\hbar}{\Delta_{a}}\sum_{l,m}{g_l g_m a^\dag_l a_m J_{i,i+1}^{lm}}\right)b_{i+1} \nonumber \\
-\frac{i}{\hbar}\left( J^{\text {cl}}+\frac{\hbar}{\Delta_{a}}\sum_{l,m}{g_l g_m a^\dag_l a_m J_{i,i-1}^{lm}}\right)b_{i-1}, \label{8b}
\end{eqnarray}
\end{subequations}
where we phenomenologically included the decay rate $\kappa_l$ of the mode $a_l$. We do not add the corresponding Langevin noise term, since we will be interested in the normal-ordered quantities only. The decay of the atoms can be included in a similar way as well, but it is usually much smaller than the cavity relaxation.

In Eq.~(\ref{8a}) for the electromagnetic fields $a_l$, the two last
terms in the parentheses in the first line correspond to the phase shift of the light
mode due to nonresonant dispersion (the second term) and due to
tunneling to neighboring sites (the third one). The second line in
Eq.~(\ref{8a}) describes scattering of all modes into $a_l$, while
the third term takes into account corrections to such scattering
associated with tunneling due to the presence of additional light
fields. In Eq.~(\ref{8b}) for the matter field operators $b_i$, the
first line gives the phase of the matter-field at the site $i$, the
second and third terms describe the coupling to neighboring sites.

It is important to underline that except for the direct coupling
between neighboring sites, as usual for the standard
Bose--Hubbard model, Eqs.~(\ref{8}) also take into account
long-range correlations between sites, which do not decrease with
the distance and are provided by the common light modes $a_l$ that
are determined by the whole set of matter-field operators $b_i$.
Such a cavity-mediated long-range interaction and nonlocal correlations between the operators $b_i$, which are introduced by the general Eqs.~(\ref{8}), can give rise to new
many-body effects beyond the standard Bose-Hubbard
model \cite{EPJD08}.

\section{QND probing of the many-body states observing light}

As the results of the previous section show, the light-matter interaction leads to the joint evolution of the light and atomic variables. The light and matter get correlated. As this is a fully quantum problem, in general, the light and matter get entangled. Thus, observing the light, one can obtain the information about the quantum states of the atoms. The goal of this section is to establish relations between the characteristics of scattered light and those of the many-body atomic system.

Considering measurements in quantum mechanics, a first goal is to obtain the expectation values of desired quantities or their distribution functions. Thus, such type of probing is intrinsically associated with multiple measurements of the same quantity and averaging over the measurement outcomes. The type of the quantum measurement considered in this section corresponds to the quantum nondemolition (QND) \cite{Brune} observation of different atomic variables by detecting light. In contrast to the completely destructive schemes usually used (e.g. time of flight measurements, absorption images), the QND measurement by light scattering only weakly perturbs the atomic sample and many consecutive measurements can be carried out with the same atoms without preparing a new sample for each measurement.

However, an important statement of quantum mechanics is, that any measurement, even a QND one, affects the quantum state of the system. Therefore, to measure the expectation values or statistics of some variable, one should prepare the system in the same quantum state before each measurement (or wait until the initial state will restor due to the system evolution). In the next section, we focus on the essentially quantum properties of the measurement process itself going beyond the simple goal of measuring only the expectation values. We will analyze, how the many-body atomic state changes during the measurement process, i.e., we will consider the measurement at a single quantum trajectory (single run of a measurement) without a requirement of the statistical averaging over the many runs of an optical experiment.

Let us first consider the model of spinless bosons introduced before to demonstrate, how the relation between the light and atom observables can be established. Then, we will provide the generalizations of that systems and measurements suggested recently. None of those examples goes beyond the typical measurements of expectation values and statistics utilizing the averaging procedure. The underlying quantum properties of the measurement process (measurement back-action at a single quantum trajectory) will be addressed in the next section.

\subsection{Model for probing bosons in optical lattices}

To focus on the question how to establish the relation between the light and atom observables, we simplify the model. We consider only two light modes: the probe $a_0$ and the scattered light $a_1$. We assume that the tunneling of atoms between the neighboring sites is much slower than light scattering, and tunneling can be neglected during light scattering. Physically this means that the quantum properties of the atomic state are determined by tunneling and interaction, but are frozen during the measurement time. In practice, the tunneling and scattering times can be different in orders of magnitude. For a deep lattice the coefficients $J_{i,i}^{lm}$ reduce to
$J_{i,i}^{lm}=u_l^*({\bf r}_i)u_m({\bf r}_i)$ neglecting atom spreading. The influence of the atom spreading within each site on the light signal can be studied even by classical scattering \cite{KetterleBragg2011}.

After those simplifications, the Hamiltonian Eqs.~(\ref{7}) takes
the form:
\begin{eqnarray}\label{1PRA09}
H=\hbar(\omega_1 + U_{11} \hat{D}_{11}) a^\dag_1 a_1+
\hbar U_{10}(\hat{D}^*_{10}a^*_0a_1 + \hat{D}_{10}a_0a^\dag_1)  \nonumber \\
-i\hbar(\eta_1^* a_1 - \eta_1 a^\dag_1),
\end{eqnarray}
where $a_1$ is the cavity-mode annihilation operator. The external probe is assumed to be in a coherent state, thus its amplitude is given by
a c-number $a_0$. $U_{lm}=g_lg_m/\Delta_a$ ($l,m=0,1$), $\eta_1$
is the amplitude of the additional probing through a mirror at the
frequency $\omega_p$ (the probe-cavity detuning is
$\Delta_{p}=\omega_{p}-\omega_1$). The operators
$\hat{D}_{lm}= \sum_{j=1}^K{u_l^*({\bf r}_j)u_m({\bf
r}_j)\hat{n}_j}$ sum contributions from all illuminated sites with
the atom-number operators $\hat{n}_j$ at the position ${\bf r}_j$.

The first term in the Hamiltonian describes the atom-induced shift of
the cavity resonance. The second one reflects scattering
(diffraction) of the probe $a_0$ into a cavity mode $a_1$. The key feature of a quantum gas is that the frequency shift and probe-cavity coupling coefficient are operators, which leads to different light scattering amplitudes for various atomic quantum states~\cite{PRL07,PRA07,NP07,LP09}.

The Hamiltonian (\ref{1PRA09}) describes QND measurements of the
variables associated with the operators $\hat{D}_{lm}$ by detecting the photon number
$a^\dag_1a_1$ or light amplitude $a_1$ related quantities (e.g. quadratures). In order for a measurement to be a QND one, several conditions should be fulfilled for the "signal observable" of interest $A_S$ ($\hat{D}_{lm}$ in this case), "probe observable" $A_P$, which is actually detected, (here, $a^\dag_1a_1$ or $a_1$) and the coupling between them through the interaction Hamiltonian \cite{Brune}. According, for example, to Ref. \cite{Brune}, the interaction Hamiltonian should be a function of $A_S$; the interaction between the signal and probe should affect the dynamics of $A_P$ ($[A_P,H]\ne 0$), whereas the signal observable should not be affected by the coupling to the probe ($[A_S,H] = 0$). In addition, measuring $A_S$, its conjugate variable is affected in an uncontrollable way, therefore the evolution should not depend on that uncontrolled variable at all. In this case, the signal observables $\hat{D}_{lm}$ depend only on the atom numbers, and their conjugate variables related to the atomic phase are not present in the Hamiltonian. One can see that all those conditions of the QND measurement are fulfilled for the Hamiltonian (\ref{1PRA09}).

Note, that one has a QND access to various
many-body variables, as $\hat{D}_{lm}$ strongly depends on the
lattice and light geometry via $u_{0,1}({\bf r})$. This is an
advantage of the lattice comparing to single- or double-well setups \cite{moore,pu,you,idziaszek,mustPRA62,mustPRA64,
javPRL,javPRA,ciracPRL,ciracPRA,saito,prataviera,javOL}.

For example, one can consider a 1D lattice of the period $d$ with
atoms trapped at $x_j=jd$ ($j=1,2,..,M$). In this case, the
geometric mode functions can be expressed as follows: $u_{0,1}({\bf
r}_j)=\exp (ijk_{0,1x}d+\phi_{0,1j})$ for traveling waves, and
$u_{0,1}({\bf r}_j)=\cos (jk_{0,1x}d +\phi_{0,1j})$ for standing
waves, where $k_{0,1x}=|{\bf k}_{0,1}|\sin\theta_{0,1}$,
$\theta_{0,1}$ are the angles between mode wave vectors ${\bf
k}_{0,1}$ and a vector normal to the lattice axis. In the plane-wave
approximation, additional phases $\phi_{0,1j}$ are $j$-independent.

For some geometries, $\hat{D}_{11}$ simply reduces to the operator
$\hat{N}_K=\sum_{j=1}^K\hat{n}_j$ of the atom number at $K$ sites
\cite{PRL07,PRA07,NP07} (if $a_1$ is a traveling wave at an arbitrary
angle to the lattice, or the standing wave with atoms trapped at the
antinodes). If the probe and cavity modes are coupled at a
diffraction maximum (Bragg angle), i.e., all atoms scatter light in
phase, $u_1^*({\bf r}_j)u_0({\bf r}_j)=1$, the probe-cavity coupling
is maximized, $\hat{D}_{10}=\hat{N}_K$. If they are coupled at a
diffraction minimum, i.e., the neighboring atoms scatter with opposite phases $0$ and $\pi$, $\hat{D}_{10}=\sum_{j=1}^K (-1)^{j+1}\hat{n}_j=\hat{N}_\text{odd}-\hat{N}_\text{even}$ is the
operator of number difference between odd and even sites. Thus, the
atom number as well as number difference can be nondestructively
measured. Note, that those are just two of many examples of how a
QND-variable can be chosen by the geometry in a many-body system.

From the Hamiltonian, the Heisenberg equation for the scattered light can be obtained as follows:

\begin{eqnarray}\label{2NatPhys}
\dot{a}_1= -i\left(\omega_1 +U_{11}\hat{D}_{11}\right)a_1
-iU_{10}\hat{D}_{10}a_0 -\kappa a_1+\eta_1,
\end{eqnarray}
where $\kappa$ is the cavity decay rate. In classical physics, such an equation is directly analogous to the Maxwell's equation for the light amplitude of the cavity mode, and the classical meaning of the $\hat{D}_{lm}$ operators (i.e. the frequency dispersion shift and coupling coefficient between two modes) is obvious. Here, in the fully quantum problem, both light- and matter-related quantities are treated as operators indeed.

The stationary solution for the operator of the light amplitude oscillating at the probe frequency takes the form
\begin{eqnarray}\label{11}
a_1=\frac{\eta_1-iU_{10} a_0\hat{D}_{10}}{i(U_{11}
\hat{D}_{11}-\Delta_p)+\kappa},
\end{eqnarray}
which gives us a direct relation between the light operator and various atom number-related operators.
It is clear that if a light-related observable is a linear function of the atom number operators $\hat{n}_j$, then the measurement of that observable will depend only on the mean atom numbers (i.e., their expectation values $\langle\hat{n}_j\rangle$). Such a measurement will carry information only about the mean atom density, which can be similar for various quantum states, and thus is not of interest for the scope of this paper. Therefore, the question is to find light observables, which depend nonlinearly on the atom number operators. In this case, the measurement will reveal the higher moments of the atom number operator, which carry information about the quantum state of ultracold atoms.

This suggests us to consider the following optical configurations, where the light observables are sensitive to various atomic states.

(I) Transverse probing \cite{PRL07,PRA07,LP09}. Here we neglect the dispersive frequency shift. In this case, the light amplitude $\langle a_1 \rangle$ is a linear function of the atom numbers and is not of great interest. However, the light intensity (given by the mean photon number $n_\Phi= \langle a_1^\dag a_1 \rangle$) already depends on the atom density-density correlations $\langle\hat{n}_i\hat{n}_j\rangle$, which differ for various atom states. Moreover, the photon number variance carries the information about the four-point correlation function $\langle\hat{n}_i\hat{n}_j\hat{n}_k\hat{n}_l\rangle$, which is even a more exciting result.

(II) The probing through a mirror, where the dispersive frequency shift plays a key role ~\cite{NP07,LP09}. In this case, even the light amplitude nonlinearly depends on the atom numbers (see the denominator in Eq.~(\ref{11})). Here, the measurement of light can directly reveal the full atom number distribution function. The measurement of frequency shifts in a cavity configuration is similar to the measurements of the phase shifts in the free-space geometry.

\begin{figure*}
\scalebox{0.9}[0.9]{\includegraphics{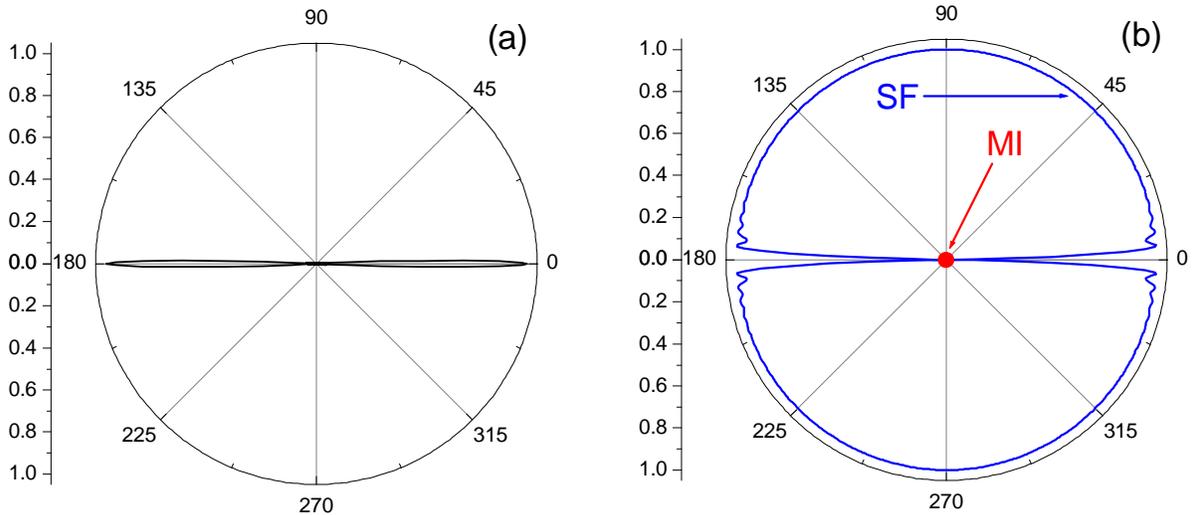}}
\caption{\label{fig2} Angular distribution of scattered light. Two traveling waves are used for probing and measurement. (a) Intensity of classical scattering shows usual diffraction peaks. (b) Quantum addition to classical scattering $R(\theta_1)$. While the Mott insulator (MI) state shows no (zero) addition, the quantum addition for scattering from the superfluid (SF) state shows the anisotropic signal proportional to the atom number $N$. $N=M=K=30$, lattice period is $d=\lambda/2$, the probe angle is $\theta_0=0$.}
\end{figure*}

Let us consider the configuration (I), where only the transverse probe $a_0$ is present (no probing through a mirror $\eta_1=0$) and the dispersive frequency shift is small (the term $U_{11}\hat{D}_{11}$ can be neglected) in Eq.~(\ref{11}). The light amplitude operator is linear in the atom number operators, $a_1=C\hat{D}_{10}=\sum_{j=1}^K A_j\hat{n}_j$, where $C=iU_{10}a_0/(i\Delta_p-\kappa)$ and $A_j=u_1^*({\bf r}_j)u_0({\bf r}_j)$. Its expectation value measures the mean atom numbers only. However, the number of photons scattered into a cavity depends quadratically on the atom operators:
\begin{eqnarray}\label{3PRL09-2}
n_\Phi= \langle a_1^\dag a_1 \rangle=|C|^2\langle\hat{D}_{10}^*\hat{D}_{10}\rangle=|C|^2\sum_{i,j=1}^K{A_i^* A_j \langle\hat{n}_i\hat{n}_j\rangle} .
\end{eqnarray}
Importantly, already a simple quantity as the mean light intensity (mean photon number $n_\Phi$) depends on the atom density-density correlations, which are second moments of the atom operators and are different for various atomic quantum states.

Similarly to the light amplitude, the quadratures of the light fields are linear in atom numbers. However, the quadrature variances depend on the density-density correlations and thus are sensitive to the atomic state. As the mean photon number is very sensitive to the second moments of the atom numbers, the photon number variance (given by the second moments of $n_\Phi$) depends on the fourth moments of the atomic operators $\langle\hat{n}_i\hat{n}_j\hat{n}_k\hat{n}_l\rangle$ and provides even more information about the atomic state through the four-point correlations. Thus, the light intensity (number of scattered photons), quadrature variance, and photon number variance all are sensitive to various atomic states with different atom number fluctuations. Here we focus only on the intensity measurements, while the quadratures and photon number variance were considered in details in ~\cite{PRA07}.

For the simplest case of two traveling light waves, the situation is almost identical to the simple diffraction in free space: the probe $a_0$ is scattered from the periodic grating of atoms and the outgoing light $a_1$ is measured at some angle to the lattice. Consequently, the angular distribution of the scattered light amplitude $\langle a_1\rangle$ shows the classical diffraction pattern. In the diffraction maxima (Bragg angles), all atoms scatter light in phase with each other (all $A_j=1)$ and the light amplitude is proportional to the mean atom number in the illuminated region $N_K$. In the diffraction minima, the mean light amplitude is zero due to the destructive interference (the neighboring atoms scatter light with the opposite phases $A_j=(-1)^{j+1}$).

However, the photon number exhibits a more interesting angular distribution. For a quantum gas, the correlation function $\langle\hat{n}_i\hat{n}_j\rangle$ generally cannot be factorized. As a consequence, the photon number is not given by the light amplitude squared, but is an essentially different quantity. For example, in the diffraction maximum the photon number is proportional to $\langle\hat{N}^2_K\rangle \ne N^2_K$. Furthermore, the light intensity in the diffraction minimum is not zero at all, but reflects the quantity $\langle(\hat{N}_\text{odd}-\hat{N}_\text{even})^2\rangle \ne 0$. The latter quantity corresponds to the fluctuations of the atom number difference between the odd and even sites. Although the atom number difference is zero in average (the light amplitude $\langle a_1\rangle$ is zero as well), its fluctuations are very different for various atomic states, which is reflected in the mean photon number $n_\Phi$.

To characterize the difference between the quantum and classical contributions to the light scattering, we introduce a quantity proportional to the difference between the photon number and the amplitude squared:
\begin{eqnarray}\label{R}
R(\theta_0, \theta_1)\equiv \langle\hat{D}^*\hat{D} \rangle - |\langle \hat{D}
\rangle|^2 = \nonumber\\
=\langle \delta\hat{n}_a\delta\hat{n}_b\rangle
|A|^2+(\langle\delta\hat{n}^2\rangle - \langle
\delta\hat{n}_a\delta\hat{n}_b\rangle)\sum_{i=1}^K{|A_i|^2},
\end{eqnarray}
which is the angle-dependent function of various atom number fluctuations. Here, $A(\theta_0,\theta_1)=
\sum_{i=1}^K{A_i(\theta_0,\theta_1)}$ is the angular distribution of classical diffraction.

The atom number fluctuations are the following: $\delta\hat{n}_i=\hat{n}_i - n$ giving the correlations between different sites $\langle\delta\hat{n}_a\delta\hat{n}_b\rangle=\langle
\hat{n}_a\hat{n}_b\rangle-n^2$, and the on-site fluctuations $\langle\delta\hat{n}^2\rangle$
equal to the variance $(\Delta n_i)^2=\langle\hat{n}_i^2\rangle-n^2$. Although the approximation that pair fluctuations are equal for all sites is not general, it hold for several interesting examples such as Mott insulator state (MI), superfluid state (SF) and the coherent state approximation to the SF state. Importantly, all those states have the same mean atom numbers $n$ at each site, but very different atom number fluctuations. We will show that such differences are readily captured by the light scattering.

In the MI state, there are no atom number fluctuations, and the light intensity shows precisely classical diffraction pattern, thus the quantum - classical difference is zero, $R_\text{MI}=0$. In the approximate coherent state, the pair correlations between the sites are neglected, $\langle\delta\hat{n}_a\delta\hat{n}_b\rangle=0$, but the on-site fluctuations are Poissonian, $\langle\delta\hat{n}^2\rangle=n$. Hence, in Eq.~(\ref{R}) the first term with classical angular distribution is zero and $R_\text{Coh}=n\sum_{i=1}^K{|A_i|^2}$. Note that the angular distribution here is completely different from the classical diffraction pattern, as the sum does not go over the light amplitudes, but rather over light amplitudes squared (i.e. the intensities of laser beams at the positions of the atoms). In the SF state, both on-site and pair fluctuations are non-zero and both two terms in Eq.~(\ref{R}) with two different angular distributions contribute to $R_\text{SF}$.

Figures 2 shows the angular distributions of two quantities: the light amplitude and quantum - classical difference ("quantum addition" for the classical signal) $R$ for the SF state. In this figure, a case of two traveling waves is presented. In classical scattering, two diffraction maxima are present, while outside them, there is a region of the destructive interference. However, the angular dependance of $R$ is different: it is nonzero at diffraction minima and zero at the directions of the diffraction maxima. Thus, the photon number is nonzero everywhere. In the diffraction maxima, it is proportional to $N^2$, while outside them it is isotropic and proportional to the atom number variance, which for the SF state is $N$. The scattering from the MI state gives zero photon number outside the diffraction maxima.

One should underline an important property of light scattering from periodic lattices, which is very convenient for experiments. Although the photon numbers in maxima and minima are different in the order in atom number, there is no problem to resolve both of them, because they are completely separated by the angles. The quantum contribution proportional to $N$ appears on top of a zero classical contribution. In the case of scattering from a homogeneous BEC, this contribution would appear on the top of the strong classical forward scattering, proportional to $N^2$, and could be hardly resolved.

A coherent product state of atoms in different wells is an approximate description of the SF state, where pair correlations are neglected and, as a consequence, the total atom number fluctuates. Although the latter fact is a definite defect of the theory, such a coherent-state approximation is usually used in the many-body theories, for example, as a step to derive the mean-field approach. It is important, that light scattering is sensitive to such an approximation and can distinguish between the SF and coherent product state. In Fig. 2, the function $R$ would show absolutely isotropic angular distribution without going to zero at the Bragg angles. Indeed, Eq.~(\ref{R}) demonstrates that $R$ is sensitive to the pair correlations and the total atom number conservation.

\begin{figure*}
\scalebox{0.9}[0.9]{\includegraphics{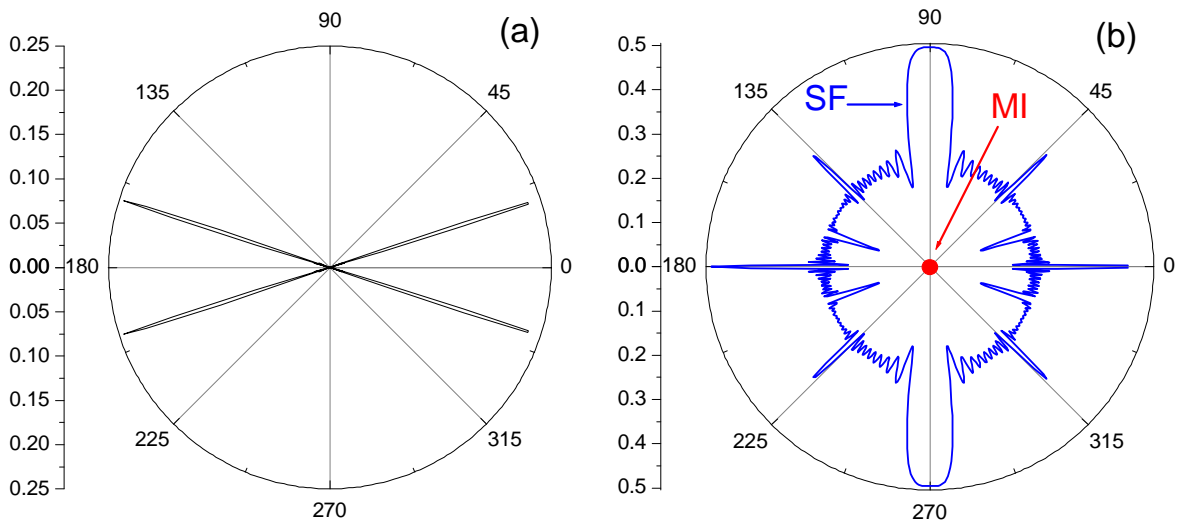}}
\caption{\label{fig3} Angular distribution of scattered light. Same as in Fig. 2, but two standing waves are used for probing and measurement. (a) Intensity of classical scattering shows usual diffraction peaks. (b) Quantum addition to classical scattering $R(\theta_1)$. While the Mott insulator (MI) state shows no (zero) addition, the quantum addition for scattering from the superfluid (SF) state shows the anisotropic signal proportional to the atom number $N$. In contrast to the traveling wave configuration (Fig. 2), the angular distribution of quantum scattering is much richer: the features appear even at the angles, where the classical diffraction does not exist. $N=M=K=30$, lattice period is $d=\lambda/2$, probe angle is $\theta_0=0.1\pi$.}
\end{figure*}

Figure 3 shows the same quantities as in Fig. 2, but for the configuration, where two standing waves are used instead of two traveling ones. The classical scattering consists of four diffraction maxima, which can be simply explained by the fact that each of two standing waves effectively consists of two traveling waves. However, the striking feature of the quantity $R$ is that its angular distribution is much richer than the classical scattering. The scattering from a quantum gas displays features at the angles, where the classical scattering does not exist at all. More precisely, in classical scattering only the zero-order diffraction maxima are possible for the parameter set considered. However, scattering from a quantum gas displays the features at angles, which would correspond to the first-order diffraction maxima (i.e. they appear at the angles corresponding to the doubled lattice period).

Such a behavior can be explained by the last term in Eq.~(\ref{R}), which sums the light intensities at the positions of the atoms. For the traveling waves, the mode functions are exponents and all $|A_i|^2=1$ giving the isotropic contribution. However, for the standing waves, the mode functions are cosines, and their squares display the period doubling and continue to depend on the position of the atoms. In Eq.~(\ref{R}), this geometrical term is multiplied by the on-site and pair fluctuations. Physically, one can say that scattering from or into a standing wave reflects not only the periodic density, but also the periodicity of the density fluctuations.

More generally, one can show that the multiple features in the fluctuation-dependent scattering appears not only for the standing waves, but for any scattering which involves the amplitude modulation inside the sample. As shown in \cite{PRA07}, similar angular features can be obtain if the quadratures are observed (even for two traveling waves), because different quadratures are coupled spatially differently to the atoms in a lattice. The angular distribution of photon statistics displays similar features as well. Another way to observe those features, and thus to probe the periodic structure of the atomic noise, is to use several probe beams, which can be traveling waves. In this case, several beams will interfere inside the sample, which results in the amplitude modulation within the lattice. The latter situation can, in principle, be realized in a recent experiment \cite{KuhrPRL} on light scattering from the ultracold atoms in optical lattices, where five probe beams were used to simultaneously cool and probe the atoms. Together with the work on light scattering from atoms in a 3D lattice \cite{KetterleBragg2011}, these are the first experiments on light scattering from ultracold atoms in optical lattices in there truly quantum regime, where the main object of interest and measurement is the light, rather than atoms in contrast to most of the works involving time-of-flight-type measurements.

Let us now turn to the configuration (II) mentioned above: probing through a mirror without any transverse probe. In Eq.~(\ref{11}), we set $a_0=0$, and the photon number operator $a^\dag_1a_1$ will be therefore represented by a lorentzian function with a frequency shift given by the atom number-dependent operator $U_{11}\hat{D}_{11}$. In the simple cases (the traveling-wave cavity or standing wave cavity with atoms trapped at their antinodes), the operator $\hat{D}_{11}$ reduces to the operator of the atom number inside the cavity $\hat{N}_{K}$. This type of measurement corresponds to the detection of cavity transmission spectrum. For a given detuning between the probe and cavity frequencies $\Delta_p$, the expectation value of the photon number is

\begin{eqnarray}\label{CavTransm1}
n_{\Phi}(\Delta_p)=\langle a^\dag_1a_1\rangle =\left\langle \frac{|\eta_1|^2}{(\Delta_p-U_{11}\hat{N}_{K})^2 +\kappa^2}\right\rangle,
\end{eqnarray}
where the expectation value is calculated for the many-body atomic state. This expression shows that the photon number depends nonlinearly on the atom number operator and, thus, is sensitive to the quantum state of atoms.

In the atomic state, where the atom number in $K$ sites does not fluctuate and takes only a single value $N_K$, the above expression is simply a classical Lorentz contour describing the cavity transmission. This is the case for the MI state. However, in the states, where the atom number fluctuates, the cavity transmission is much more complicated. As was shown in \cite{NP07} for the good cavity limit, the transmission spectrum consists of multiple lorentzians forming a comb-like structure (see Fig. 4 for the comparison of MI and SF states). The peaks correspond to all possible atom numbers in a cavity, thus, the peaks appear at all possible dispersion frequency shifts. Moreover, as was proved in \cite{NP07}, such a transmission spectrum directly maps out the full atom number distribution function. Thus, by measuring the transmission spectrum of a cavity containing ultracold atoms in some particular quantum state, one can directly obtain the full atom number distribution function corresponding to that quantum state.

\begin{figure*}
\scalebox{1.7}[1.7]{\includegraphics{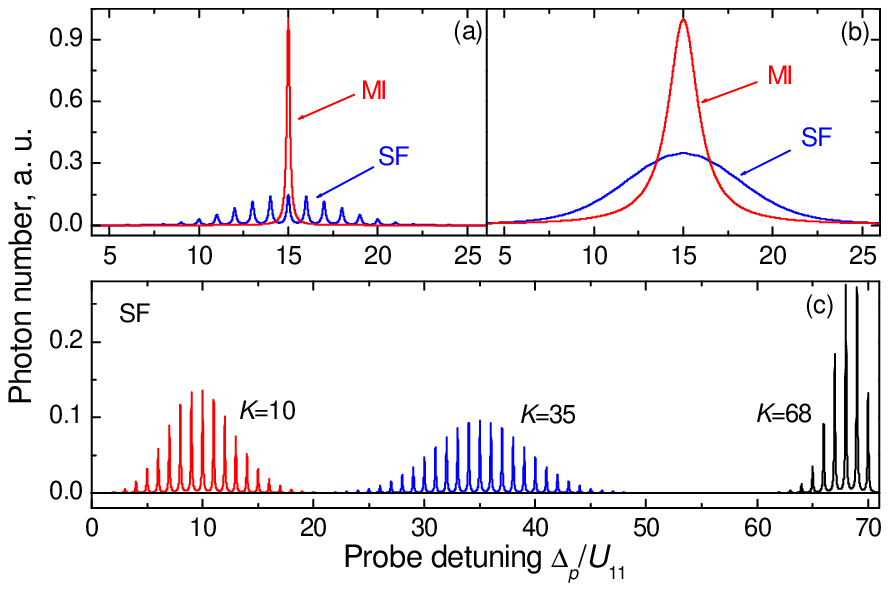}}
\caption{\label{fig4} Transmission spectra of a cavity. The spectra directly map out the full atom number distribution functions of an ultracold gas. (a) Single Lorentzian for MI reflects the non-fluctuating atom number.
Many Lorentzians for SF reflect the atom number fluctuations,
which are imprinted on the positions of narrow resonances in the spectrum. Here, the cavity is good and all satellites are resolved (the cavity decay rate $\kappa$ is smaller than the satellite separation $U_{11}$, $\kappa=0.1U_{11}$). $N=M=30$, $K=15$. (b) The same as in (a) but the cavity is worse ($\kappa=U_{11}$), which gives smooth broadened contour for SF. Although the satellites are not resolved, the spectra for SF and MI states are very different. (c) Spectra for SF with $N=M=70$ and different number of sites illuminated $K=10,35,68$. The transmission spectra have different forms, since different atom distribution functions correspond to different $K$. $\kappa=0.05U_{11}$.}
\end{figure*}

Figure 4 shows the cavity transmission spectra (and hence directly the form of the atom number distribution) for MI and SF states. The atom number fluctuations are completely suppressed in MI state and one sees only a single classical Lorentz contour. In SF state, the atom number distribution is close to the Poissonian one (and hence close to the Gaussian one for some conditions). In Fig. 4(c), the transmission spectra are shown for different number of sites $K$ illuminated. One observes how the transmission spectrum and the atom number distribution changes depending on $K$. For example, the distribution is narrow, when almost all sites are illuminated, because the total atom number is fixed and the fluctuations are suppressed. In contrast, when a smaller part is illuminated, the fluctuations are rather large and the distribution function is broad.

If the separated peaks cannot be resolved due to the large cavity linewidth, the comb-like structure smooths, but its width and form still give the information about the atom number distribution function (Fig. 4(b)). Interestingly, in the superfluid state one can get a convolution of the Lorentz and Gauss functions, which is the so-called Voigt contour, well known in physics of hot atomic ensembles with Doppler broadening. Here, it does not originates from the motion of hot atoms, but due to the quantum uncertainty in the position of ultracold atoms \cite{NP07}. Changing the optical geometry, one can get access to the full distribution functions of various atom number-related quantities \cite{NP07}, e.g., to that of $N_\text{odd}-N_\text{even}$.

Let us summarize the results of this section, which established the relations between the properties of the quantum sates of atoms and light observables using the example of bosonic atoms in optical lattices. The measurement of scattered light distinguishes between various quantum states of ultracold atoms even if their mean densities are the same. Light scattering is sensitive to the global atom fluctuations (i.e. in the whole lattice or its significant part), local fluctuations (related to the variables at a single site) and distant correlations between two or more lattice sites (e.g. the four-point correlations). Surprisingly, the interesting observation angle is not the Bragg one, but the direction of a diffraction minimum, where the strong classical scattering is suppressed and the light carries information about the quantum fluctuations directly. We demonstrated how the Mott insulator state, superfluid state, and the approximate coherent state all can be distinguished by light scattering. The use of standing-waves or several probe beams (alternatively, the quadrature measurements) provides us the information about the spatial structure of the atomic noise. Moreover, the full atom number distribution functions of various atomic variables can be obtained by measuring spectral (or other phase-dependent) properties of the scattered light.

\subsection{Other models and systems: finite temperatures, fermions, spins, molecules}

 In the previous section we presented the detailed results for the case, where the atomic tunneling between neighboring sites was assumed slow, and atomic motion was frozen during the measurement. The influence of tunneling was explicitly taken into account in the general model in Sec. II. From the general Hamiltonian Eq.~(\ref{7}) and Heisenberg equation, Eq.~(\ref{8}), it is evident that the dynamics of light and atomic operators are essentially coupled to each other and the long-range interaction length can greatly exceed the distance between neighboring sites \cite{PRA07}. Atomic tunneling affects the phase and amplitude of scattered light, and the light scattering modifies the tunneling process (e.g. it affects the phase and amplitude of the atomic operators). The modification of the light scattering for the time intervals, where tunneling is important, was considered, e.g., in \cite{PhysRevA_77_033620LiuTRansmissionTunneling, PhysRevA.81.013404MorigiLikeWePlusTunneling}. In Ref. \cite{PhysRevA_77_033620LiuTRansmissionTunneling}, the short- and long-scale time regimes we identified for the case of a BEC in a double-well potential. In Ref. \cite{PhysRevA.81.013404MorigiLikeWePlusTunneling}, the influence of the photon-induced atomic recoil on the site-to-site hopping was analyzed. A question of the light dynamics was addressed in Ref. \cite{PhysRevA_75_023812MeystreFirst}.

Here, we consider bosons at zero temperature. A theoretical model of light scattering from bosonic atoms taking into account the finite temperature effects was developed in Ref. \cite{PhysRevA_80_043404IdziaszekFiniteTnoCavity}.

Ultracold gases are not limited to the bosonic atoms and models for the interaction of quantum light with ultracold matter can be generalized for various involved many-body states of fermions, spin particles, ultracold molecules, etc. For example, the thermometry of fermions in optical lattices based on light scattering was suggested in Refs. \cite{PhysRevLett_103_170404RuostekoskiFermions,Burnett2011}, where the idea of the light detection outside the diffraction maxima was shown useful as well.

The consideration of fermions and spins links this direction of research to the many-body models of condensed matter physics even stronger. Currently, in the ultracold gas community, there is a significant interest in such many-body phenomena as quantum magnetism, fermionic superfluidity, etc. \cite{KetterleFermionReview}. Quantum light scattering is a powerful tool to get insight in those effects, understand their properties and origins, which still have not been fully clarified, and measure their quantum characteristic.

Considering the ultracold gas of spin fermions, one should underline an important analogy to the system of bosons in optical lattices presented above. In the system considered here, the phase of scattered light essentially depends on the position of an atom and the detection angle. Such a dependence made possible to obtain the constructive interference (at Bragg angles) and completely destructive interference (at diffraction minima). More generally, the variation of the observation angle was used as a method to tune the relative phase of light waves scattered from different atoms. For the spin particles, the atomic levels corresponding to different spins can be usually split by some frequency interval and one can use the spin-selective light scattering for probing those states. Different polarizations and frequencies of light couple differently to various spin levels. The phase difference of light scattered from different spin components can be tuned by changing the frequency of probing light. Depending on the detuning between the probe frequency and two spin levels, on get constructive or, importantly, destructive interference between the light waves scattered from two spin components. Changing the probe frequency one can indeed go continuously from the totaly constructive to totally destructive interference, which is a direct analogy of the diffraction maxima and diffraction minima considered above. One of the important consequences of such a method, is that at some frequency, the mean light amplitude can vanish (destructive interference), and the photon number will directly reflect the fluctuations of the population difference between two spin levels, thus, revealing the pairing between two spin states. In this example, the measurement of the fluctuations of the atom number difference between two levels $N_\uparrow - N_\downarrow$ directly corresponds to the measurement of the atom number difference between odd and even sites $N_\text{odd}-N_\text{even}$ in a diffraction minimum presented earlier in this paper. As the strong classical scattering is suppressed in both cases, the light intensity directly measures the fluctuations of those differences given by $(\hat{N}_\uparrow - \hat{N}_\downarrow)^2$ or $(\hat{N}_\text{odd}-\hat{N}_\text{even})^2$. Creation of spin pairs (Cooper pairs) corresponds to the suppressed spin difference fluctuations, while for the unpaired states the fluctuations are large. As in the case of model presented above, this is just one example of the important quantity, which can be measured. Changing the frequencies or angle of light, one can select different variables (as the operators $\hat{D}_{lm}$) to be measured in a QND way.

The pairing of spins into Cooper pairs is a very important phenomenon responsible for the appearance of the states such as Bardeen-Cooper-Schrieffer (BCS) or Fulde-Ferrell-Larkin-Ovchinnikov (FFLO) phases. Several theoretical models have been suggested to characterize those and other many-body phases of spin particles (including quantum magnetism) by light scattering. The proposals involving the QND measurements of spin statistics beyond the mean-field approximation include Refs. \cite{PolzikPRL,PolzikNaturePh,DemlerQND,PolzikNJP,PolzikArXiv2011,Cirac2008,deChiaraPRA2011}. Analogously to the bosons in optical lattices, the idea to use the standing wave light for measuring the spatial structure of the atomic noise was confirmed in Refs. \cite{PolzikNaturePh,PolzikNJP} for spins. As shown in Ref. \cite{PolzikNaturePh},  various quantum antiferromagnetic states could be characterized by the polarization-sensitive methods of light scattering. While most experimental works on quantum spin fluids are currently focused on the mean-field characteristics (e.g. using the doubling of the spatial period of spins when going from paramagnetic/ferromagnetic to antiferromagnetic phase) \cite{HuletNature2010,HuletPRA}, the experiments \cite{KetterlePRL2011,KetterleSpinsArXiv2011} successfully reports the measurements of spin fluctuations observed by scattered light.

Similarly to the case of bosons considered above, the cavity transmission spectroscopy can be used to obtain the distribution functions of the spin variables in different sates of quantum magnets \cite{PhysRevA_79_013630ChinaLikeWe,PhysRevA_80_043623Liu,OptExpr2010RamanLiu,KZhangPRA2011}.

Recently, the method of light scattering was applied to analyze different systems: ultracold dipolar molecules \cite{PRL2011Harvard,PRA2011Harvard,LP12} trapped in low dimensions (two one-dimensional tubes). One finds that in such a system the formation of the few-body molecular complexes is possible. In particular, the work goes beyond the dimers (similar to those appearing in the BCS-BEC theories with two paired spins), but also predicts the existence of trimers and tetramers (the bound sates of three and four polar molecules, respectively). The scheme of the light detection at a diffraction minimum was generalized and proved effective for those few-body complexes. As in the example of bosons in an optical lattice considered here, the light scattering is sensitive to the position of the particles. It was therefore shown that it is sensitive to the relative distances between the molecules forming the complex. The dissociation (or association) of the complex is accompanied by the increase (or decrease, respectively) of the particle number fluctuations, which can be readily observed in the intensity of scattered light. This opens a way for the optical nondestructive in situ detection of the quantum many-body states of ultracold molecules and their dynamics in real time.

\section{Single-run measurements and measurement-based preparation of the quantum states}

\begin{figure*}
\scalebox{1.5}[1.5]{\includegraphics{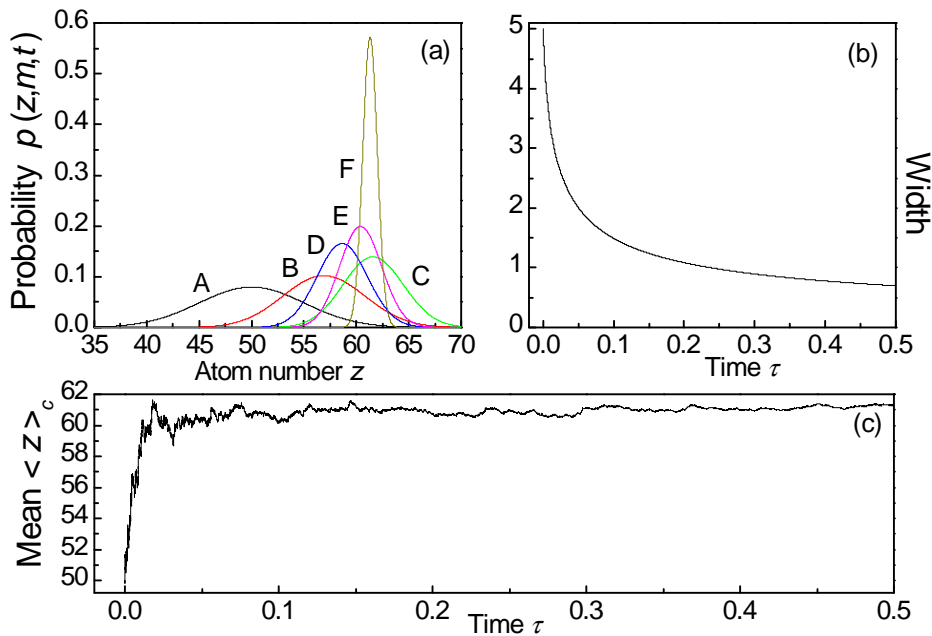}}
\caption{\label{fig5} Photodetection at diffraction
maximum leading to the atom number squeezed state. Results for a single quantum trajectory (Quantum Monte Carlo simulations). (a) Shrinking atom number distribution at different times $\tau=$ 0, 0.005, 0.018, 0.03, 0.05, 0.5 (A-F). (b) Decreasing width of the distribution $\delta z$. (c) Mean atom number $\langle z\rangle_c$ stabilizes after many quantum jumps. Initial state: SF, $N=100$ atoms, $K=M/2=50$ illuminated sites.}
\end{figure*}

In all problems considered above, the physical mechanism of light scattering establishes a relation between the light and matter observables. As we have shown, the light provides the information about various statistical quantities of the quantum states of atoms: different correlation functions (given by the expectation values of some operators) and distribution functions of different variables can be measured. As it is usual in quantum mechanics, the determination of such statistical quantities requires multiple measurements. Therefore, the repeated preparation of the initial state is necessary, because any quantum measurement (even a QND one) generally affects the quantum state of the system.

In this section, we will address the problem from another point of view going beyond the standard goal of measuring the expectation values and distribution functions. Here, we focus on a system dynamics during a single run of the optical measurement (i.e. the continuous detection of scattered photons), without taking the average (expectation values) over many realizations as we were doing so far \cite{PRL09,PRA09,LP10,LP11}. Indeed, the result of a single-run measurement is important, as it is the first result one obtains in an experiment before the averaging procedure.

During the interaction, the light and matter get entangled. According to quantum mechanics, due to the entanglement, the measurement of one of the quantum subsystems (light) will also affect the other quantum subsystem (atoms), as an example of the quantum measurement back-action. While measurement back-action often is an unwanted perturbation, we will use it to change the atomic state in a desired way. We will consider the light measurement as an active method to prepare particular many-body quantum states of the ultracold atoms.

The Hamiltonian of the system is given by Eq.~(\ref{1PRA09}). Even though some dynamics due the atom tunneling has been neglected here, we will show that there is still nontrivial dynamics (quantum jumps and non-unitary evolution) exclusively associated with the quantum measurement process. The initial motional state of the ultracold atoms trapped in the
periodic lattice potential at the time moment $t=0$ can be
represented as

\begin{eqnarray}\label{2PRA09}
|\Psi^a(0)\rangle =\sum_{q}c_q^0 |q_1,..,q_M\rangle,
\end{eqnarray}
which is a quantum superposition of the Fock states corresponding to
all possible classical configurations $q=\{q_1,..,q_M\}$ of $N$
atoms at $M$ sites, where $q_j$ is the atom number at the site $j$.
For each classical configuration $q$, the total atom number is
conserved: $\sum_j^M q_j=N$. This superposition displays the
uncertainty principle, stating that at ultralow temperatures even a
single atom can be delocalized in space, i.e., there is a
probability to find an atom at any lattice site. We will show, how
this atomic uncertainty is influenced by the light detection.

For example, for a limiting case of the MI state, where the atom
numbers at each lattice site are precisely known, only one Fock
state will exist in Eq.~(\ref{2PRA09}): $|\Psi_\text{MI}\rangle
=|1,1,..,1\rangle$ for the MI with one atom at each site. On the
other hand, the SF state is given by the superposition of all
possible classical configurations with multinomial coefficients:

\begin{eqnarray}\label{3PRA09}
|\Psi^a_\text{SF}\rangle=\frac{1}{(\sqrt{M})^N}\sum_{q}
\sqrt{\frac{N!}{q_1!q_2!...q_M!}}|q_1,q_2,..q_M\rangle.
\end{eqnarray}
Thus, the atom number at a single site as well as the atom number at
$K<M$ sites are uncertain in the SF state.

We use the open system approach \cite{Carmichael} to describe the
continuous counting of the photons leaking out the cavity. When the photon is detected at the moment $t_i$, the quantum jump occurs, and
the state instantaneously changes to a new one obtained by applying
the cavity photon annihilation operator $|\Psi_c(t_i)\rangle
\rightarrow a_1|\Psi_c(t_i)\rangle$ and renormalization (the
subscript $c$ underlines that we deal with the state conditioned on
the photocount event). Between the photocounts, the system evolves
with a non-Hermitian Hamiltonian $H-i\hbar\kappa a^\dag_1a_1$. Such
an evolution gives a quantum trajectory for $|\Psi_c(t)\rangle$
conditioned on the detection of photons at times $t_1,t_2,...$.

Importantly, due to several assumptions and simplification, one can get a transparent analytical solution of the problem. First, after the tunneling is neglected, the Hamiltonian does not mix the Fock states in the
expression (\ref{2PRA09}). So, the problem is reduced to
finding solutions for each atomic configuration
$q=\{q_1,..,q_M\}$. The full solution will then be given by
the superposition of those solutions.

Second, it is known~\cite{Denis,GardinerZoller} that, if a coherent probe
illuminates a prescribed atomic configuration in a cavity, the light
state is proportional to a coherent state $|\alpha_q(t)\rangle$ with
$\alpha_q(t)$ given by Maxwell's equations. Each atomic Fock state in Eq.~(\ref{2PRA09}) will then be correlated with a coherent light state with parameters given only by
the corresponding configuration $q$: $|\Psi_c(t)\rangle
=\sum_{q}c_q^0\exp[\Phi_q(t)]
|q_1,...,q_M\rangle|\alpha_q(t)\rangle/F(t)$, where $F(t)$ gives the
normalization. So, the problem to find $|\Psi_c(t)\rangle$ reduces
to finding $\alpha_q(t)$, $\Phi_q(t)$ for all classical
configurations forming the initial state $|\Psi(0)\rangle$.

The third mathematical simplification arises due to the fact that for coherent states of light, the quantum jumps do not lead to a discontinuity of the light amplitude evolution $\alpha_q(t)$ \cite{PRA09}. Moreover, after $t_1 \gg 1/\kappa$ a steady state is achieved for all $\alpha_q(t)$. Interestingly, in contrast to many problems in quantum optics, where the
steady state is a result of averaging over many quantum trajectories,
here, the steady state in $\alpha_q(t)$ appears even at a single quantum trajectory. This is a particular property of the coherent state of light with damping.

Although a solution is available for any $t$ (cf. Refs. \cite{PRA09,Meystre09}), we present it for $t>1/\kappa$, when the steady state is achieved in all $\alpha_q(t)$, and assuming
the first photon was detected at $t_1>1/\kappa$.

Due to the steady state in all $\alpha_q(t)$, the solution is
independent of the detection times and after $m$ counts is
\begin{eqnarray}
|\Psi_c(m,t)\rangle =\frac{1}{F(t)}\sum_{q}\alpha_q^m e^{\Phi_q(t)}
c_q^0 |q_1,...,q_M\rangle|\alpha_q\rangle, \label{2PRL09}\\
\alpha_q=\frac{\eta-iU_{10} a_0D^q_{10}}{i(U_{11}
D^q_{11}-\Delta_p)+\kappa}, \label{3PRL09}\\
\Phi_q(t)=-|\alpha_q|^2\kappa t+(\eta\alpha^*_q-iU_{10}
a_0D^q_{10}\alpha^*_q-\text{c.c.})t/2, \label{4PRL09}
\end{eqnarray}
where $D^q_{lm}= \sum_{j=1}^K{u_l^*({\bf r}_j)u_m({\bf r}_j)q_j}$ is
a realization of the operator $\hat{D}_{lm}$ at the classical configuration $\{q_1,..q_M\}$; $a_0$, $\eta$,
and $\alpha_q$ all oscillating in steady state at $\omega_p$ were
replaced by their constant amplitudes.

In contrast to a single atomic Fock state, the full solution,
in general, is not factorizable into a product of the atomic and
light states, and the light and matter are entangled.
Moreover, in contrast to a single Fock state, the quantum jump
(applying $a_1$) changes the state, and the evolution of the full
$|\Psi_c(m,t)\rangle$ is not continuous. Note that even for $t>1/\kappa$,
when all $\alpha_q$ reached their steady states and are constants,
this solution is still time-dependent. Thus, the time
$t=1/\kappa$ is not characteristic for the steady state
of the full solution.

As we see, each light amplitude $\alpha_q(t)$, Eq.~(\ref{3PRL09}), is
given by a Lorentzian corresponding to classical optics.
Eq.~(\ref{2PRL09}) shows that the probability to find an atom
configuration $q$,
$p_q(m,t)=|\alpha_q|^{2m}\exp{(-2|\alpha_q|^{2}\kappa
t)}|c_q^0|^2/F^2$, changes in time due to the photodetection. This
demonstrates the back-action of the light measurement on the atomic
state and the essentially quantum measurement-induced dynamics of the system, not related to any obvious processes as tunneling, etc.

In the following, we will show consequences of Eq.~(\ref{2PRL09}) for two
cases (I) and (II) addressed in the previous section, where only one probe ($a_0$ or $\eta$) exists. As before, for the transverse
probing ($a_0\ne 0)$, we also neglect the mode shift, assuming
$U_{11} D^q_{11}\ll\kappa$ or $\Delta_p$. Thus, in both examples,
$\alpha_q$ (\ref{3PRL09}) depends on the configuration $q$ only via a
single statistical quantity now called $z$: $z=D^q_{11}$ for cavity
probing ($\eta \ne 0$), and $z=D^q_{10}$ for transverse probing.

From Eq.~(\ref{2PRL09}) we can determine the probability distribution of
finding a given $z$ after time $t$ as
\begin{eqnarray}\label{5PRL09}
p(z,m,t)=|\alpha_z|^{2m}e^{-2|\alpha_z|^2\kappa t}p_0(z)/F^2,
\end{eqnarray}
where the initial distribution $p_0(z)=\sum_{q'} |c_{q'}^0|^2$, such
that all configurations $q'$ have the same $z$; $F^2= \sum_z
|\alpha_z|^{2m}\exp{(-2|\alpha_z|^2\kappa t)}p_0(z)$ provides
normalization.

The solution gets especially simple for the case of a macroscopic BEC with a large atom number. This situation was analyzed in Ref. \cite{LP10}.

\subsection{Measurement-induced number squeezing}

As was mentioned,
at the Bragg angle, $\hat{D}_{10}=\hat{N}_K$ is the operator of atom
number at $K$ sites. So, $z$ varies from $0$ to $N$ reflecting
possibilities to find any atom number at $K$ sites. The light
amplitudes (\ref{3PRL09}) $\alpha_z=Cz$ are proportional to $z$,
$C=iU_{10} a_0/(i\Delta_p-\kappa)$. The probability (\ref{5PRL09}) reads
\begin{eqnarray}\label{6PRL09}
p(z,m,t)=z^{2m}e^{-z^2\tau}p_0(z)/\tilde{F}^2
\end{eqnarray}
with a characteristic time $\tau=2|C|^2\kappa t$.

When time progresses, both $m$ and $\tau$ increase with a
probabilistic relation between them. The Quantum Monte Carlo method
\cite{Carmichael} establishes such a relation, thus giving a
trajectory. Note, that thanks to the simple analytical solution
(\ref{2PRL09}), it gets extremely simple. In each step, it consists in
the calculation of the photon number in the state given by
Eq.~(\ref{2PRL09}) and comparing it with a random number generated in
advance, thus, deciding whether the detection or no-count process
has happened.

If the initial atom number $z$ at $K$ sites is uncertain, $p_0(z)$
is broad [for the superfluid (SF) it is nearly Gaussian
\cite{NP07,PRL09,PRA09,LP10}], and Eq.~(\ref{6PRL09}) shows that $p(z,m,t)$ is
strongly modified during the measurement. The function
$z^{2m}\exp{(-z^2\tau)}$ has its maximum at $z_1=\sqrt{m/\tau}$ and
full width at half maximum (FWHM) $\delta z \approx
\sqrt{2\ln2/\tau}$ (for $\delta z \ll z_1$). Thus, multiplying
$p_0(z)$ by this function will shrink the distribution $p(z,m,t)$ to
a narrow peak at $z_1$ with the width decreasing in time
(Fig. 5).

This describes the projection of the atomic quantum state to a final
state with squeezed atom number at $K$ sites (a Fock states
$|z_1,N-z_1\rangle$ with $z_1$ atoms at $K$ sites and $N-z_1$ atoms
at $M-K$ sites). Surprisingly, when $\delta z<1$, the final collapse is even
faster than $\sqrt{\tau}$, due to the discreteness of $p(z,m,t)$ \cite{LP11}.
Measuring the photon number $m$ and time $t$, one can determine
$z_1$ of a quantum trajectory.

In contrast to recent results in spin squeezing,
which can be also obtained for thermal atoms
\cite{PolzikHot,HollandNJP2008}, in our work, quantum nature of ultracold
atoms is crucial, as we deal with the atom number fluctuations
appearing due to the delocalization of ultracold atoms in space.

After the distribution shrinks to a single $z_1$, the light
collapses to a single coherent state $|\alpha_{z_1}\rangle$, and the
atoms and light get disentangled with a factorized state
\begin{eqnarray}\label{7PRL09}
|\Psi_c\rangle=|z_1,N-z_1\rangle|\alpha_{z_1}\rangle.
\end{eqnarray}
So, light statistics evolves from super-Poissonian to Poissonian.
The conditioned (i.e., at a single trajectory) cavity photon number
$\langle a^\dag_1a_1\rangle_c(t)=|C|^2 \sum_{z=0}^Nz^2p(z,m,t)$ is
given by the second moment of $p(z,m,t)$. Its dynamics [very similar
to $\langle z\rangle_c$ in Fig. 5(c)] has jumps, even
though all $\alpha_{z}(t)$ are continuous. In the no-count process,
$\langle a^\dag_1a_1\rangle_c$ decreases, while at one-count it
jumps upwards, which is a counter-intuitive signature of super-Poissonian statistics and conditional probabilities.
Finally, it reduces to $\langle a^\dag_1a_1\rangle_c=|C|^2z_1^2$,
reflecting a direct correspondence between the final atom number and
cavity photon number, which is useful for experiments.

Importantly, even the final many-body Fock state still contains atom-atom entanglement,
as many components $|q_1,..,q_M\rangle$ can have the same $z_1$. For
example, the SF state can be represented as $|SF\rangle_{N,M}=\sum_z
\sqrt{B_z} |SF\rangle_{z,K}|SF\rangle_{N-z,M-K}$ ($B_z$ are binomial
coefficients). After the measurement, it ends up in
$|SF\rangle_{z_1,K}|SF\rangle_{N-z_1,M-K}$, i.e., the product of two
uncorrelated superfluids.

Our measurement scheme determines (by squeezing) the atom number at
a particular lattice region and projects the initial atomic state to
some subspace. However, the atom number at different regions keeps
quantum uncertainty. So, the quantum structure of the final state
can be revealed in a subsequent optical or matter-wave experiment.
Thanks to the lattice geometry, one can change the illuminated
region, and further study measurement-induced collapse of the state
in the remaining subspace.

Even in matter-wave experiments \cite{BlochSFMI}, the product of
SFs will look different from the initial SF: the atoms from
different regions will not interfere in average. Note, that we did
not specify how $K$ sites were selected. One can illuminate a
continuous region. However, one can illuminate each second site by
choosing the probe wavelength twice as lattice period and get number
squeezing at odd and even sites. In this way, one gets a
measurement-prepared product of two SFs ``loaded'' at sites one by
one (e.g. atoms at odd sites belong to one SF, while at even sites
to another). While the initial SF shows the long-range coherence
$\langle b^\dag_i b_j\rangle$ with the lattice period, the prepared
state will demonstrate the doubled period in $\langle b^\dag_i
b_j\rangle$ ($b_j$ is the atom annihilation operator).

This example shows, that the phase of the matter field can be manipulated by this type of QND measurements, even though we measure only the occupation number-related operators. This is a direct consequence of the many-body nature of the atomic state. As usual in the QND measurements, by measuring the number-related variables, one typically destroys the conjugate variable (in this case, the phase). However, here we have a very rich choice of the number-related variables for the measurement. Measuring some of the variables, other variables can be left untouched. Therefore, one can carefully destroy only particular phase information in our system, while other phase-related variables stay unaffected by the measurement. This is demonstrated in the last example, where the phase coherence survives after the measurement, but changes its period.

\subsection{Schr{\"o}dinger cat state preparation}

\begin{figure*}
\scalebox{1.5}[1.5]{\includegraphics{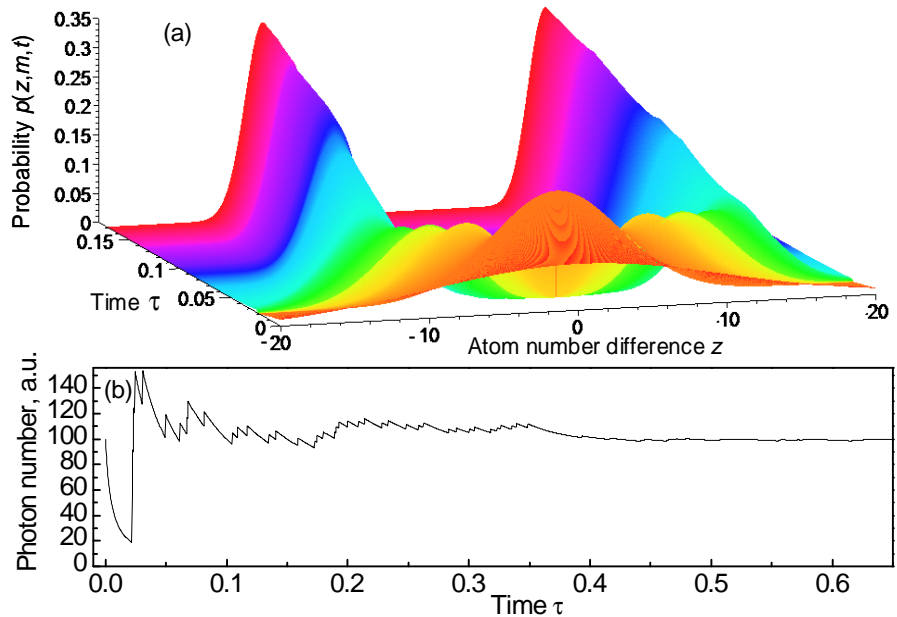}}
\caption{\label{fig6} Photodetection at diffraction minimum leading to the generation of a Schr{\"o}dinger cat state. Results for a single quantum trajectory (Quantum Monte Carlo simulations). (a) Shrinking  distribution of the atom-number difference between odd and even sites for various times. The doublet corresponds to Schr{\"o}dinger cat state. (b) Conditional photon number with quantum jumps. Initial state: SF, $N=100$ atoms, $K=M=100$ sites.}
\end{figure*}

In contrast to
classical atoms, quantum gases scatter some light even into diffraction
minima \cite{PRL07,PRA07,NP07,LP09}. Here $z=D^q_{10}=\sum_{j=1}^M
(-1)^{j+1}q_j$ is the occupation number difference between odd and even
sites, varying from $-N$ to $N$ with a step 2 (we assumed $K=M$).
Eq.~(\ref{6PRL09}) keeps its form with a new meaning of $z$ and $p_0(z)$
[for SF, new $p_0(z)$ is nearly a Gaussian centered at $z=0$ and the
width $\sqrt{N}$ \cite{NP07,PRL09,PRA09,LP10}].

The striking difference to the diffraction maximum is that our
measurement (\ref{6PRL09}) is not sensitive to the sign of $z$, while the
amplitudes $\alpha_z=Cz$ are. So, the final state is a macroscopic
superposition of two Fock states with $z_{1,2}=\pm \sqrt{m/\tau}$
and light amplitudes: $\alpha_{z_2}=-\alpha_{z_1}$,
\begin{eqnarray}\label{8PRL09}
|\Psi_c\rangle=(|z_1\rangle|\alpha_{z_1}\rangle+
(-1)^m|-z_1\rangle|-\alpha_{z_1}\rangle)/\sqrt{2}.
\end{eqnarray}
Figure 6 shows the collapse to a doublet probability $p(\pm
z_{1},m,t)$ and the photon-number trajectory, where upward jumps and
no-count decreases can be seen.

In contrast to a diffraction maximum, even in the final state, light and
matter are not disentangled. However, to keep the purity of the
state, one should know the number of detected photons,
because of the sign flip in Eq.~(\ref{8PRL09}). This reflects the
fragility of macroscopic superposition states with respect to the
decoherence.

The atom number squeezed states prepared by observing
light at a diffraction maximum are much more robust than the
Schr{\"o}dinger cat state obtained at a diffraction
minimum to photon lost and detection efficiency, as the former do not have any phase jump associated with the photocounts. However, the
convenient property of the measurement at a minimum is that during
the same time interval the number of photons scattered at a diffraction
minimum ($\langle a_1^\dag a_1\rangle =|C|^2N$) is much smaller than
the one scattered at a maximum ($\langle a_1^\dag a_1\rangle
=|C|^2N_K^2$) \cite{PRL07,PRA07,PRL09,PRA09}. Thus the cat state is not rapidly destroyed by the decoherence as one may expect from the coherent Bragg scattering picture.

Various types of photon statics during the measurement-based state preparation were considered in Ref. \cite{PRA09}.

\subsection{Quantum state preparation by the cavity-transmission measurements}

Let us now consider the prototype configuration (II) from the previous section: transmission probing through a cavity mirror. In the previous section, we have shown that the transmission spectrum has a comb-like structure and maps out the full atom number distribution function. Such a spectrum is demonstrated in Fig. 4. As discussed before, this spectrum reflects the result of multiple measurements. The question now is what one can get in a single run of the optical measurement. Naively, one can expect that in each run one gets one of the peaks in Fig. 4, which would correspond to the collapse to the single-peak distribution function. Only after averaging over many runs one would recover the full comb-like spectrum in Fig. 4. The situation is however much more interesting in details. Depending on the realization (i.e. for different quantum trajectories), the distribution function after a single-run of the transmission spectrum measurement can consist either of one peak or of two peaks. A one-peak distribution function corresponds to the atom number squeezing, while a two-peak function signals the generation of the Schr{\"o}dinger cat state.

Thus, probing through a mirror represents an interesting example, where either a number squeezed state or a Schr{\"o}dinger cat state is prepared depending on the measurement outcome. Knowing the properties of the system after the measurement, one can understand, which kind of the states was prepared. Moreover, the cat state obtained using this configuration has a more general form than in the case of the detection at the diffraction minimum: the amplitudes of two cat components can be non-equal and the phase difference between them can take various values (not restricted to $\pi$ as in the diffraction minimum). The latter property makes the preparation of the cat states more robust to the decoherence and photon loss \cite{PRA09}.

The full analytical expressions for the probabilities of such type of quantum measurement are complicated and were reported in Ref. \cite{PRA09}. If for a particular quantum trajectory the number of photocounts is
larger than some critical value, the distribution $p(z,m,t)$
collapses to a single-peak function centered at $z_p$. This corresponds to the collapse to a Fock state $|z_p,N-z_p\rangle$ with precisely known atom number $z_p$ within $K$ lattice sites inside a cavity. If, however, the number of photocounts is small, the
distribution $p(z,m,t)$ collapses to a doublet centered at $z_p$
with two satellites at $z_{1,2}=z_p\pm\Delta z$. This corresponds to the generation of the Schr{\"o}dinger cat state with generally imbalanced amplitudes and various phase shifts between the two components:

\begin{eqnarray}\label{27PRA09}
|\Psi_c\rangle=\frac{1}{F'}[e^{im\varphi +
i\Phi(t)}|z_1,N-z_1\rangle|\alpha_{z_1}\rangle \sqrt{p_0(z_1)}  \nonumber\\
+ e^{-im\varphi - i\Phi(t)}|z_2,N-z_2\rangle|\alpha_{z_2}\rangle
\sqrt{p_0(z_2)}],
\end{eqnarray}
where the phases are $\varphi=-\arctan({U_{11}\Delta z /\kappa})$ and $\Phi(t)=|\alpha_{z_1}|^2 U_{11}\Delta z t$.

Physically, tuning the probe frequency to $\Delta_p$, we may expect maximum scattering
for an atom number $z_p$ providing such a dispersive frequency shift
$\Delta_p=U_{11}z_p$. If the actual photocount number is large,
indeed, the atom number is around $z_p$ and it collapses to this
value. However, if $m$ is small, we gain knowledge that the atom
number $z$ is inconsistent with this choice of $\Delta_p$, but two
possibilities $z<z_p$ or $z>z_p$ are indistinguishable. This
collapses the state to a superposition of two Fock states with
$z_{1,2}$, symmetrically placed around $z_p$.

The preparation of the particular state is a probabilistic process. However, importantly, varying the detuning $\Delta_p$ between the probe frequency and the cavity resonance, one can change the distribution of possible outcomes and increase or decrease the probability of a certain state to appear. In such a way, one can, e.g., increase the probability of the cat state with a given imbalance of the amplitudes of two components. One can always chose the detuning in such a way that the amplitudes will be equal to each other independently of the measurement outcome.

In this section, we demonstrated that the cavity enhanced light scattering off an ultracold gas in an optical lattice constitutes a quantum measurement with a controllable form
of the measurement back-action. Moreover, the
class of emerging atomic many-body states can be chosen via the
optical geometry and light frequencies, as an eigenstate of the operator $\hat{D}_{lm}$. The ability to implement those operators and, thus, to design the measurement back-action is a truly advantage of the optical lattice geometry in contrast to the simpler setups with a BEC in a single- or double-well potential \cite{PRL12a,PRL12b,PRL12c,PRL12d,Denis2010}, where the flexibility is much more restricted. Moreover, the use of a cavity selects the prevailing scattering direction and assures the controllable state collapse, rather than complete destroying of the initial atomic state as it is the case for the spontaneous emission into all angles \cite{DaleyPRA2010}. Such a measurement-based state preparation is an example of the close connection of the measurement and decoherence (i.e. the photon leakage and measurement) in quantum physics \cite{ZollerPRA,ZollerNP}.

Considering the projection as a tool for state preparation, one can prepare various partitions of the systems. In the examples above, one prepared the bipartite systems: atoms in the region of $K$ sites and outside that region, atoms in odd and even sites. More complicated partitions are also possible. For example, the light scattering from a limited lattice region and detection at a diffraction minimum will prepare a tripartite system: the atoms can be in odd or even sites, within or outside of the lattice region. The initial state of such a system can be expressed via the trinomial coefficients. Therefore, using light scattering, one can simplify the initial problem of numerous atoms and sites to the simpler problems of just several subsystems. Importantly, the manipulation of particular variables can leave other variables untouched by the light scattering, allowing the consecutive operations on the same sample in the standard QND sense.

Note, that the quantum state preparation is probabilistic. However,
it can be better controlled by including a feedback loop, which will
enable ones the quasi-deterministic state preparation. In this case,
the trapping potential should be continuously modified depending on
the outcome of the photodetector. For example, the detection of
photons at a diffraction maximum squeezes the atomic number at $K$
sites around some value $z_1$, which was not known a priori. The
potential can be continuously tilted in a way to provide the
increase or decrease of this atom number to enable ones to obtained
the number squeezed state with a mean value $\tilde{z}_1$ given a
priori. The same method can be applied for the atom number squeezing
at odd or even sites.

\section{Quantum optical lattices}

\begin{figure}
\scalebox{0.8}[0.8]{\includegraphics{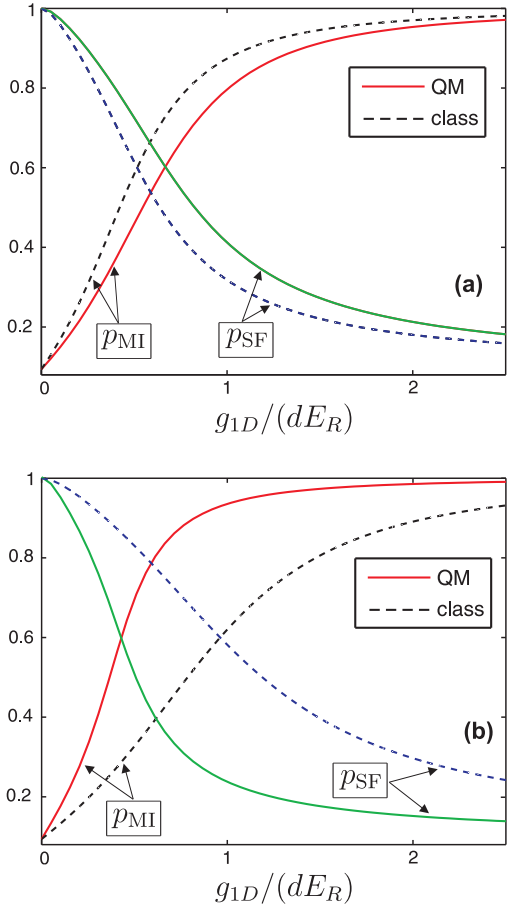}}
\caption{\label{fig7} MI to SF phase transition in a quantum optical lattice. Cavity influence on the transition is demonstrated by the comparison of the occupation probabilities $p_{\textrm{MI}}$ and $p_{\textrm{SF}}$ (to find the states $|\Psi_\text{MI}\rangle$ and $|\Psi_\text{SF}\rangle$) for a purely quantum field, i.e., $V_{cl}=0$, and a purely classical field, i.e., $\eta_1=0$, as a function of the dimensionless 1D on-site interaction strength $g_{1D}/(dE_R)$ ($d$ is the lattice period, $E_R$ is the recoil energy). We choose $\eta_1$ such that both potentials are of equivalent depth, $V=5.5E_R$, for zero on-site interaction ($g_{\textrm{1D}}=0$). The quantum (QM) and classical (class) cases are depicted with solid and dashed lines, respectively. $(U_{11},\kappa,\eta_1)=(-1,1/\sqrt{2},\sqrt{5.5})\omega_R$, where the recoil frequency is $\omega_R=E_R/\hbar$. The detuning between the probe and cavity frequencies (counted from the mean dispersive frequency shift) affects the position of the phase transition. In (a) this detuning is positive $\Delta_p-U_{11}N=\kappa$ and the transition point is shifted towards higher interaction strengths in comparison to that in a classical lattice. In (b) the detuning is negative $\Delta_p-U_{11}N=-\kappa$ and the transition point is shifted towards the smaller interaction strengths.}
\end{figure}

In the previous sections the atoms were assumed to be trapped in a prescribed potential formed by stationary laser light, which can be described by a c-number function. We have shown, how additional light probes can measure or modify the properties of the atomic quantum state due to the light-matter entanglement. In those examples, the quantum natures of both the atomic motion and probe light were important, whereas the potential was treated classically.

In this section, we address the more general regime, where the quantum nature and fluctuations  of the trapping light potential itself play a key role \cite{PRL05,EPJD08}. Indeed, the potential is created by light, which is a quantum object. The goal is to consider the systems and phenomena, where the quantum nature of the potential cannot be neglected and the trapping potential is a quantum dynamical variable. It must be determined self-consistently with the solution for the quantum states of atoms trapped in that potential.

The results of the previous sections can be conceptually considered as a perturbation theory for the ultimate regime of fully quantum optical lattices. If the trapping beams are rather strong, the potential can be represented by its mean value and additional quantum fluctuations. Then, the quantum fluctuations will obey the equations very similar to those for the quantum probe light mode considered before. In this way, the entanglement and mutual dynamics of light and matter fluctuations can be described, including the measurement back-action. Such a perturbative regime with small quantum fluctuations of the trapping potential can be realized even with optical lattices in free space (without a cavity). In this section, however, we would like to go beyond such a perturbation theory and consider the ultracold atoms trapped inside a cavity, where the potential is fully quantum.

A natural implementation to study ultracold quantum gas in quantized light is to load the atoms into a high-Q cavity. In this case, the light mode of the cavity (e.g., a single standing wave, in the simplest case) will form the trapping potential. The quantum properties of the light mode become more important, when the light intensity (the number of photons in a cavity) gets smaller and smaller. The relative value of quantum fluctuations then increase strongly, because the mean light amplitude can decrease down to zero. The striking property of the cavity configuration is that, although the light intensity decreases almost to zero, the depth of the optical potential in a cavity can be still kept very large. Indeed, the potential depth is given by the product of the photon number and the light-matter coupling coefficient. For example, the quantum potential formed by a single standing wave mode $a_1$ with the wave vector $k$ is given by $\hbar U_{11}a^\dag_1a_1\cos^2{kx}=\hbar (g^2_{1}/\Delta_a)a^\dag_1a_1\cos^2{kx}$, where $g_1$ is the light-matter coupling coefficient and $\Delta_a$ is the frequency detuning between the light and atomic resonance (cf. Eq.~(\ref{3}) and the discussion after it). In a cavity, the light-atom coupling coefficient can reach huge values. Thus, even the light field of a single photon can lead to a very deep optical potential inside a cavity. The trapping and cooling of atoms by a single photon inside a cavity has been already demonstrated experimentally \cite{Rempe} (the atom in those works was not however cold enough for its motion being quantized).

As we have shown in this paper before, the quantum states of atoms and light strongly depend on each other. For example, the presence of atoms shifts the cavity resonance and modifies the light amplitude because of light scattering. On the other hand, the atoms are trapped in that light field, which requires a self-consistent solution for the coupled light-matter quantum dynamics. Importantly, quantum mechanics allows the superpositions of several Fock states of photons forming the quantum potential. Therefore, one can consider a superposition of potentials of several depths. This rises an intriguing question about the possibility to have the superposition of several atomic phases, each of which is correlated to different Fock state of light (i.e. to the potentials of different depths). For example, one can imagine the potential provided by a very week cavity mode, where the Fock states of zero photons (the vacuum field), one photon and two photons are significant. Then, in principle, one can have a superposition, of the free particles (correlated with the zero photon number Fock state), the superfluid state (correlated with a potential provided by a single photon) and Mott insulator state (correlated with the two-photon Fock state). Of course, the stability of such an intriguing superposition and its robustness with respect to the decoherence should be carefully analyzed.

Another remarkable property of the quantum optical lattices in a cavity is the existence of the long-range interaction between the atoms beyond the standard Hubbard models. As demonstrated by the Hamiltonian (\ref{7}) and Heisenberg equations (\ref{8}), even if the tunneling is non-negligible only between the neighboring sites (as it is usual for the Bose-Hubbard model), even very distant atoms can still interact with each other via the common cavity light mode. As a consequence, the tunneling of very distant atoms gets correlated (sometimes, called co-tunneling). Mathematically, this is expressed by the coefficients in the generalized Bose-Hubbard model, which are not constants, but the operators depending on the whole set of light and atomic operators in the whole extended optical lattice. Thus, the cavity mediates the long-range interaction, whose properties can be tuned by the cavity parameters. Even for the simplest type of the neighboring interaction, the parameters of the Bose-Hubbard model influencing tunneling, atom-atom interaction and atom localization can be controlled by the parameters such as probe-cavity detuning, probe intensity, probe-atom detuning, cavity linewidth, etc. \cite{EPJD08}.

As a prominent example, the quantum phase transition between the MI and SF states in a fully quantum optical lattice was analyzed numerically in a finite system in Ref. \cite{EPJD08}. Fig. 7 compares the phase transitions in quantum and classical optical lattices of the same mean potential depths. From the classical point of view, the lattices of the same depth assure the same physics. However, in quantum lattices, the potential depth is not the only important parameter. In particular, near the phase transition point the photon number fluctuations play a key role. More precisely, if the transition occurs for the potential depth given by the mean photon number $n_{\Phi}$, already the photon numbers $n_{\Phi}\pm 1$ are associated with different atomic phases. Thus the photon fluctuations drive the atomic fluctuations and hence the phase transition. Depending on the cavity parameters (e.g. the cavity frequency), the photon fluctuation can either suppress or enhance the atomic fluctuations and atomic hopping, therefore driving the system towards or outwards of the MI or SF states. In Fig. 7 one sees that the position of the phase transition in a cavity can be shifted towards either smaller or larger values of the atom-atom interaction strengths, depending on the detuning between the probe light and the cavity resonance. (The probe-cavity detuning counted from the mean dispersive shift of the cavity resonance is positive in Fig 7(a) and negative in Fig 7(b)). This demonstrates the tunability of the phase transition properties by the cavity parameters.

\begin{figure}
\scalebox{0.55}[0.55]{\includegraphics{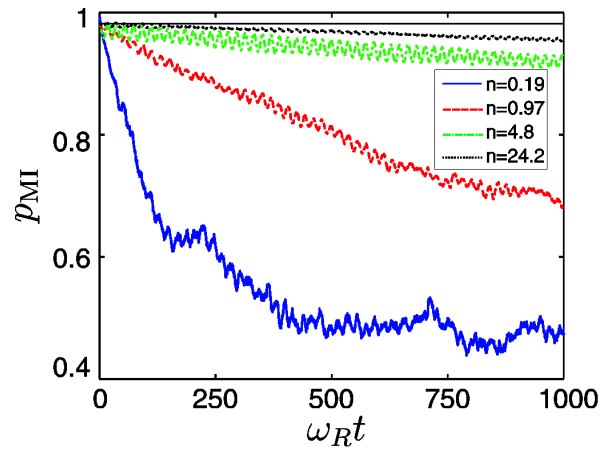}}
\caption{\label{fig8} Dynamics of the quantum state in a quantum potential. Probability $p_\text{MI}$ to find the MI state $|\Psi_\text{MI}\rangle$ for various photon numbers forming the quantum potential. In all cases the potential depth is the same ($8E_R$). Classically, the same potential depth would mean the same physics. However, in a quantum optical lattice, the evolution for the same depth, but different photon numbers is strikingly different. For the classical potential (constant line) the system stays in the MI state without any change. Decreasing the photon number in a cavity (the mean photon numbers are 24.2, 4.8, 0.97 and 0.19), the role of light fluctuations increases, which drives the atomic fluctuations destroying the MI state. The smaller the mean photon number, the stronger the systems evolution deviates from the case of a classical optical potential \cite{EPJD08}.}
\end{figure}

Figure 8 compares dynamics of the atomic states in classical and quantum lattices of various photon numbers. Importantly, the mean potential depths are chosen to be the same, so classically one should not expect any difference between all those curves. In contrast, one observes a striking difference in the evolutions of the probability to find the system in the Mott insulator state for different cavity photon numbers. First, for the classical potential there is no evolution and the system which started in the MI remains in it. For large photon numbers, the MI state gets destroyed with time, but does not significantly deviate from the classical solution (because the quantum fluctuations are much smaller than the mean photon numbers). However, for small photon numbers, the fluctuations are larger than the mean value and the system leaves the MI state completely. This is another example of how the photon fluctuations induce the atom hopping and atom fluctuations and radically change the quantum state of the system. Fluctuations towards lower photon number enhance tunneling and destroy the Mott phase.

The tunneling dynamics in quantum optical lattices was analyzed in \cite{EPJD08,AndrasNJP,Wolfgang2010,LarsonPRA2010}. A fundamental difference of the quantum lattices from their classical analogues is that the coupling of atoms to the leaking light mode opens a new dissipation channel for atomic dynamics. In quantum simulations with classical optical lattices, the main advantage of the ultracold atoms in contrast to the condensed matter systems is their almost complete isolation from the environment. Using the quantum optical lattices one can make a step further. The decoherence can be introduced in the atomic system in a very careful and controllable way: as we have shown in the previous section, the decoherence (the measurement back-action) can be tailored using the optical geometry. The results of Refs. \cite{EPJD08,AndrasNJP} show that the photon number fluctuations induce the relaxation of the atom oscillation between the sites. More interestingly, looking at a single quantum trajectory, the tunneling can be completely stopped, when the photon is emitted from the cavity, projecting the system to the dark state of the tunneling operator. The next photon escape restores the tunneling of the atoms.

A possibility to generate non-classical states of light using the quantum optical lattices at the single-photon level was predicted in \cite{Wolfgang} using the few-photon induced nonlinearities.

The light field in a cavity can show the bistable behavior. Therefore, the trapping potential can be bistable as well and different atomic phases can correspond to different potential depths. A very exotic phase diagram for the MI-SF phase transition in a cavity was obtained in Refs. \cite{LewenstPRL,LewenstNJP}. It was shown that the different MI insulator lobes in the phase diagram (corresponding to the different filling factors) can overlap with each other, which is not the case for ultracold atoms in a free space optical lattice. The phase diagram significantly depends on the cavity parameters.

\section{Variety of phenomena in the "ultracold gas + optical cavity" system}

Recent experimental success in trapping a BEC inside a cavity \cite{Brennecke,Colombe,Slama} has stimulated a significant attention to this system from both experimental and theoretical sides. Such a system has already shown a rich behavior for the coupled dynamics of the cavity light and atomic motion. As not all of those effects require a fully quantum description and can be explained using mean-field approaches either for light or atoms, the detailed review of those phenomena is beyond the scope of the present papers. However, we will mention some of them here, because a deep analysis of the variables, which can carry the information about the essentially quantum properties, can lead to the determination of particular features available only in truly quantum systems (e.g. the light-matter entanglement).

One of the interesting aspects of the "BEC + cavity" system is its analogy to the systems known in the field of optomechanics \cite{EsslingerScience2008}. Standard optomechanics considers the coupling between the light and mechanical oscillators \cite{Marquardt}. Typically, the setup includes the cavity with one movable mirror, where the light mode is coupled to the vibrations of that mirror. This simple model was extended to various cavities and complicated mechanical structures, where the light can be trapped.

Recent interest in optomechanics was focused on the reaching the quantum regime, where the mechanical oscillator should be cooled down to very low temperatures, which is very demanding experimentally. As it was shown, a BEC in a cavity represents a very similar system, where the motion of the mechanical oscillator is replaced by the motion of the BEC. Indeed, the motion of an ultracold atomic ensemble is one of the examples of the macroscopic mechanical oscillators. Moreover, the BEC is already in its ground state, so the quantum regime for such an ultracold oscillator is already reached. Therefore, besides representing a significant interest on its own, the BEC in a cavity can serve as a system for modeling processes in conventional optomechanics, including its quantum regime.

In such novel optomechanical systems, the bistability and nonlinear dynamics have been already demonstrated experimentally for the few-photon light in a cavity \cite{EsslingerScience2008,EsslApplPhys}. The theories of such effects, including the quantum treatment, were developed in \cite{EPJD09Domokos,PhysRevA.81.043639Domokos, PhysRevLett_102_080401DomokosNoiseDeplitionPLUSoptomech,DeChiaraPRA2011}. Mean-field approaches for various extensions of that systems (e.g. the combined system of a BEC and movable mirror in the same setup) were considered theoretically in \cite{PhysRevA_79_033401LIU,PhysRevA_80_043607Bhattach, PhysRevA.81.053833Meystre,PhysRevA_78_043618LIU2wellMeanfield,JPhysB2010,OptCommun2008BhatSFinCavity, ZubairyPRA2011,LiuarXiv11033577}. In the combined system, the bistable behavior of the MI to SF phase transition was addressed \cite{PhysRevA_80_011801MeystreBistableMIFI, OptCommun2010MeystreBistableMISF}. The motion of fermions inside a cavity was analyzed in \cite{PhysRevLett.104.063601MeystreOptomFermions,PhysRevA.83.043606Liu1DFermionsinCavity}. Various effects, analogous to the optomechanical ones can arise in the systems of spins inside a cavity \cite{Vuletic,StamperK,ZhouPRL2009,ZhouPRA2010}.

A very interesting phenomenon appears, when the atoms are illuminated by the light beam transverse to the cavity axis. If the light intensity exceeds some threshold value, the atoms self-organize in a periodic checkerboard pattern occupying only all odd or even sites of the optical lattice (the optical lattice is itself a result of the interference between the cavity and transverse light modes). Such a self-organization phenomenon was proved to be linked to the famous Dicke phase transition and the supersolid quantum phase. Using the BEC, this effect was obtained experimentally in \cite{nature09009Esslinger,EsslingerArXiv2011}, and its properties were analyzed theoretically in Refs. \cite{EPJD2008NagyMeanField, EPJDDomokos2011, PRL2010Domokos, PhysRevA.81.043407MorigiPhases, PhysRevLett.105.043001SimonsPlusOptomech,MustArxiv}. The many-body aspects of such a system were underlined in Refs. \cite{GoldbartNP,GoldbartPRA,SachdevPRL11}.

In this paper we have considered only the far off-resonant interaction of light with particles. The main attention was hence paid to the coupling between the light an atomic motion (i.e. the external degrees of freedom). Tuning the light closer to the resonance will involve the atomic internal excitations. If the ultracold gas is not destroyed by the spontaneous emission, this will enrich the system as at least three quantum oscillators will play a significant role. The consideration of atomic excitations brings this field closer to the research on excitons and polaritons in optical cavities (see, e.g., Refs. \cite{PhysRevA.81.023617SimonsExcPlusPhaseTr, PhysRevA_78_023634ChinaDickePhaseTrans, HashemPhysRevA.79.023411,HashemPhysRevA.80.053608,HashemPhysRevA.83.063831}).

\section{Conclusions}

Quantum optics of ultracold quantum gases is a quite recently developed research direction, which treats light and ultracold particle motion at full quantum level. Early theoretical proposals, first considered as a kind of Gedankenexperiments only, triggered an active and fast experimental development, which theories now hardly can follow. Particular attention in this field now is paid on addressing systems and phenomena, where the use of a fully quantum approach is really unavoidable to get the physical insight, while the standard mean-field descriptions of light or matter fail. One should address the specific quantities and system properties that carry information about the essentially quantum features of the physical phenomena.

Development of this direction can link various fields of atomic, molecular, optical and condensed matter physics, and even generate the models and approaches for other areas of theoretical and experimental physics. For example, the methods for probing intriguing quantum states already discussed can be applied for other systems and effects as well: various topological states of matter, Dirac fermions \cite{NewJPhys2010LewensteinFermions}, Kondo physics \cite{PhysRevB.81.245314KondoDetectionOfComplicated}, detection of states in fermionic \cite{annurev-conmatphys-070909-104059tbp} and superconducting systems \cite{PhysRevB_76_174519DetectionSupercond}, etc. The methods can be also applied for development of laser cooling schemes \cite{njp9_5_055025SalzburgerDifMaxMin} and precision measurement techniques \cite{PhysRevA_80.063834HindsPrecMeas,njp8_7_073014HollandSpinSqueezing, PhysRevA_79_023841MolmerSpinQND, PhysRevA_80_043803HollandprobeBlochosc}. The system of quantum gas trapped in a cavity provides a significant extension for setups considered as candidates for quantum simulators (i.e. atoms in classical optical lattices). First, the use of quantum optical lattices will enable to simulate the broader range of trial Hamiltonians. Second, the dissipation can be carefully introduced, tailored and controlled in such a system using optical methods. The multipartite entangled states of light and many-body atomic system, which naturally appear at this ultimate quantum level of the light-matter interaction, can serve as a resource for quantum information processing.

\begin{acknowledgments}
This work was supported by the EPSRC (project EP/I004394/1) and the Austrian Science Fund, FWF project  F4013.
\end{acknowledgments}

\end{document}